\documentclass{elsart}

\usepackage{graphicx}

\usepackage{amssymb}

\begin{document}

\begin{flushright}
LA-UR-06-1881
\end{flushright}
\begin{frontmatter}



\title{High Voltage Test Apparatus for a Neutron EDM Experiment and
  Lower Limit on the Dielectric Strength of Liquid Helium at Large Volumes}


\author{J. C. Long \corauthref{cor1}}
\address{LANSCE-NS, Los Alamos National Laboratory, Los Alamos
  NM 87545 USA}
\author{P. D. Barnes, J. G. Boissevain, D. J. Clark, M. D. Cooper}
\author{J. J. Gomez, S. K. Lamoreaux, R. E. Mischke, S. I. Penttila}
\address{Physics Division, Los Alamos National Laboratory, Los Alamos
  NM 87545 USA}
\corauth[cor1]{Corresponding author. Present address: Indiana
  University Cyclotron Facility, Bloomington IN 47408 USA}

\begin{abstract}
A new search for a permanent electric dipole moment (EDM) of the neutron
is underway using ultracold neutrons produced and held in a bath of
superfluid helium.  Attaining the target sensitivity requires maintaining
an electric field of several tens of kilovolts per centimeter across the
experimental cell, which is nominally 7.5~cm wide and will contain about 4
liters of superfluid.  The electrical properties of liquid helium are
expected to be sufficient to meet the design goals, but little is known
about these properties for volumes and electrode spacings appropriate to
the EDM experiment.  Furthermore, direct application of the necessary
voltages from an external source to the experimental test cell is
impractical.  An apparatus to amplify voltages in the liquid helium
environment and to test the electrical properties of the liquid for
large volumes and electrode spacings has been constructed.  The device
consists of a large-area parallel plate capacitor immersed in a 200
liter liquid helium dewar.  Preliminary results show the breakdown strength of
normal state liquid helium is at least 90~kV~cm$^{-1}$ at these
volumes, at the helium vapor pressure corresponding to 4.38~K.  These
fields hold for more than 11 hours with leakage currents less than
170~pA (about 20\% of the maximum tolerable in the EDM experiment).
The system is also found to be robust against anticipated radiation
backgrounds.  Preliminary results for superfluid show that fields of at least
30~kV~cm$^{-1}$ can be sustained at the volumes required for the EDM
experiment, about 60\% of the design goal.  These results are likely
limited by the low pressure that must be maintained above the superfluid bath.
\end{abstract}
\end{frontmatter}

\section{Introduction}
\label{sec:intro}
The search for a permanent electric dipole moment (EDM) of the neutron
has been the subject of experimental investigations for nearly 50
years \cite{Smith57}.  It has generally been of interest as a test of
the discrete space-time symmetries, and has more recently emerged as
one of the most sensitive tests of physics beyond the Standard 
Model \cite{Khriplovich97}. 

A nonzero EDM ($d_{e}$) would be a signal of time-reversal symmetry violation
and has yet to be observed.  The most sensitive upper limit to date,
$d_{e}\leq 6.3 \times 10^{-26}$~e-cm (90\% C. L.), is derived from experiments
performed on ultracold neutrons (UCN) \cite{Harris99}. UCN have
kinetic energies of a few hundred nanoelectron volts---less than the Fermi
potential of many substances---and are easily trapped.  The neutron
velocities in these experiments are small and randomized, 
which is very advantageous for the suppression of systematic effects 
associated with motional magnetic fields.  A new EDM search, with
potentially unprecedented sensitivity, is underway using UCN produced via the 
superthermal process (the downscattering of cold neutrons by phonons)
in a superfluid helium bath and subsequently held
in the bath \cite{Khriplovich97,Golub94,EDMCollaboration02}.

The statistical sensitivity 
$\sigma_{d}$ of these experiments can be approximated by the expression:
\begin{equation}
\sigma_{d}\geq \frac{\hbar}{E \sqrt{TNt}}
\end{equation}
where $\hbar$ is Planck's constant, $E$ is the electric field to which
the neutron sample is exposed, $T$ is the experimental
measurement cycle time, $N$ is the number of neutrons sampled per
cycle, and $t$ is the cumulative storage time.  For the experiment under
construction, the target neutron sample is $N \sim$~$1 \times
10^{6}$, using neutrons downscattered from a dedicated beam at the 
Spallation Neutron Source (SNS) under construction at the Oak Ridge National 
Laboratory.  The target sample time is $T \sim$ 500~s, which is
expected to be determined primarily by the beta decay lifetime of the
UCNs held in the very inert environment of the superfluid bath.  A
total of about $2\times 10^{5}$ measurement cycles are expected, for a
cumulative storage time of $\sim 10^{8}$~s.  The target electric field
is 50~kV~cm$^{-1}$.

The dielectric properties of liquid helium (LHe) are expected to be
sufficient to meet the electric field design parameter, however, very
little is known about these properties for volumes and electrode
spacings appropriate to the experiment.  The reference design calls
for a measurement cell with a volume of about 4 liters of superfluid (SF)
LHe, positioned between a set of parallel-plate high voltage (HV)
electrodes of nominal area $A_{e} = 2000$~cm$^{2}$.  The electrode
spacing and hence the cell width is nominally $z_{e} = 7.5$~cm.

\begin{figure}[htbp]
\begin{center}
\includegraphics[width=12cm]{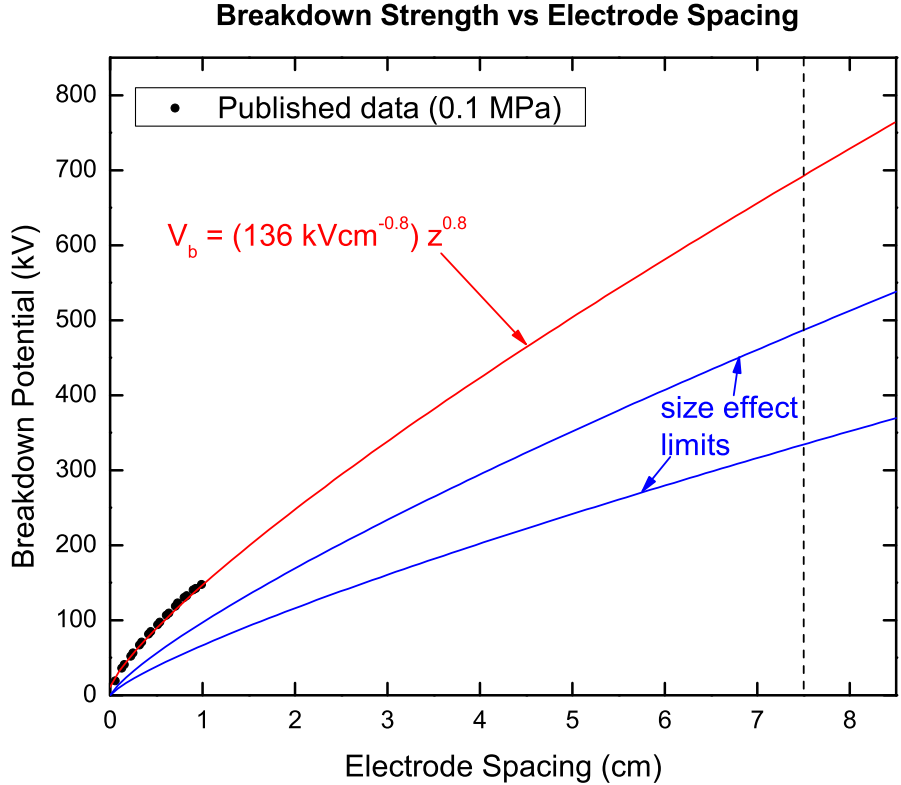}
\caption{\label{fig:projections} Breakdown voltage vs. electrode
  spacing for saturated LHe.  The black dots are existing experimental
  data from \cite{Gerhold98_1}.  The extrapolation suggested in
  \cite{Gerhold98_2} is shown, together with the further reduction
  (``size effect limits'') expected in the EDM experiment due to the larger
  volume of LHe relative to that in \cite{Gerhold98_1}.}
\vspace{1cm}
\end{center}
\end{figure}

The existing data for the electric breakdown strength of LHe
are summarized in Fig.~\ref{fig:projections}, which is adapted 
from \cite{Gerhold98_2}.  At electrode spacings below 1~mm, breakdown
fields well in excess of 
300~kV~cm$^{-1}$ are observed.  This reduces to 150~kV~cm$^{-1}$ for
gaps of 1~cm, however.  Furthermore, as of the inception of the project
described in this work, reliable data for
larger spacings were not available in the literature.  The
extrapolation of the breakdown voltage to larger spacings 
suggested in \cite{Gerhold98_2} is $V_{b} = Cz^{0.8}$, where $V_{b}$ is
in kV~cm$^{-1}$, $z$ is in cm and $C = 136$ kV~cm$^{-0.8}$.  This
extrapolation is illustrated in Fig.~\ref{fig:projections}, and
suggests an expected breakdown field of about 95~kV~cm$^{-1}$ at the
nominal electrode spacing.  This is well above the design target, if an
extrapolation of nearly an order of magnitude in the electrode spacing
is to be trusted.  

As emphasized in \cite{Gerhold98_2}, however,
electric breakdown in cryogenic liquids is influenced by many factors,
including the volume of stressed liquid.  In particular, the expression
\begin{equation}
E_{b2}=E_{b1}\left(\frac{S_{2}}{S_{1}}\right)^{-1/m}
\label{eq:size_effect}
\end{equation}
is cited in \cite{Gerhold98_2} as an order-of-magnitude estimate for
the reduction in breakdown field $E_{b2}$ as a function of the overall
stressed volume $S_{2}$, relative to reference values $E_{b1}$ and
$S_{1}$.  Suggested values of the exponent $1/m$ range in the
literature from about 0.08 to 0.17.  According to \cite{Gerhold98_1}, 
the data in Fig.~\ref{fig:projections} are obtained from an experiment with
spherical electrodes of radius $r = 3.1$~cm.  Assuming circular plate
electrodes of the same radius for simplicity and the sake of an 
order-of-magnitude estimate, the stressed volume
$S_{1}$ is given by $\pi r^{2}z$, where $z$ is the separation.
Substituting this and the relation $S_{2} =
A_{e}z$ for the stressed volume in the EDM experiment into
Eq.~\ref{eq:size_effect} yields an estimate for the
expected reduced breakdown field between the EDM electrodes relative
to the experiment in \cite{Gerhold98_1}.  As the breakdown
voltage scales with the field for the parallel plate electrodes,
scaling the extrapolated curve in Fig.~\ref{fig:projections} by
$(S_{2}/S_{1})^{-1/m}$ yields the plots 
labeled ``size effect limits'' in that figure.  This is the modified
expectation for the breakdown voltage in the EDM experiment.  The 
upper and lower bounds correspond to the exponent $1/m =$ 0.08 and
0.17, respectively, suggesting a wide possible range in the vicinity of
the design voltage (about 375~kV for the 7.5~cm electrode gap).

Additional factors influencing breakdown cited in \cite{Gerhold98_2}
include surface roughness, stressing time, and liquid pressure, for 
which understanding and predictability are partial at
best.  Furthermore, the foregoing data and discussion apply only to saturated
LHe in the normal state.  Evidence presented in \cite{Gerhold98_2}
suggests additional reduction in breakdown strength for the SF state 
relevant to the EDM experiment.  This state of affairs motivates a 
direct experimental assessment of the dielectric properties
of LHe at large volumes and electrode spacings, and in the SF state.

In addition to the breakdown strength of LHe at large volumes,
several other engineering parameters are of interest to the
neutron EDM experiment.  Of greater interest than the pure breakdown 
strength is a stable operating point, defined as the maximum voltage 
attainable with a breakdown probability of at most a few percent over 
the entire anticipated integration time of the experiment ($\sim
10^{8}$~s).  This value will ostensibly be chosen as the experimental 
operating voltage in order to protect the EDM detection system, which
will be SQUID--based and likely damaged in the event of a spark
discharge.  The electric field stability over time $\Delta E_{t}$ is also of
interest, with a target of $\Delta E_{t} \leq 1$\% required over the 
course of a measurement cycle to suppress false EDM signals generated by
stray magnetic fields \cite{EDMCollaboration02}.  This translates into
a leakage current upper limit of about 1~nA for the EDM HV system.
The electric field uniformity $\Delta E_{V}$ is of similar importance,
with a target of $\Delta E_{V} \leq 1$\% required over the
experimental cell volume to further control systematics
associated with motional magnetic fields \cite{EDMCollaboration02,Lamoreaux96}.

Finally, direct application of the required 
voltages from an external power source would require electrical
feedthroughs impractically large for the cryogenic environment.
Therefore, an apparatus has been designed to test a simple in-situ
amplification mechanism in addition to testing the parameters listed
above.  In this mechanism, a small initial potential is used to charge
a variable, parallel--plate capacitor in the LHe volume.  The potential 
source is then disconnected and the voltage amplified by separating
the plates and reducing the capacitance.  In the reference design of
the final version of the EDM experiment, the amplifier (variable)
capacitor is connected in parallel to a pair of fixed capacitors (with the
nominal electrode dimensions mentioned above) that flank the actual
measurement cells \cite{EDMCollaboration02}.  In the test system
described here, only the
variable capacitor has been prototyped, which should be sufficient to
test the dielectric properties of LHe and to verify the amplification
principle.

\section{Apparatus}
\subsection{Central Volume}
\begin{figure}[htbp]
\begin{center}
\includegraphics[width=12cm]{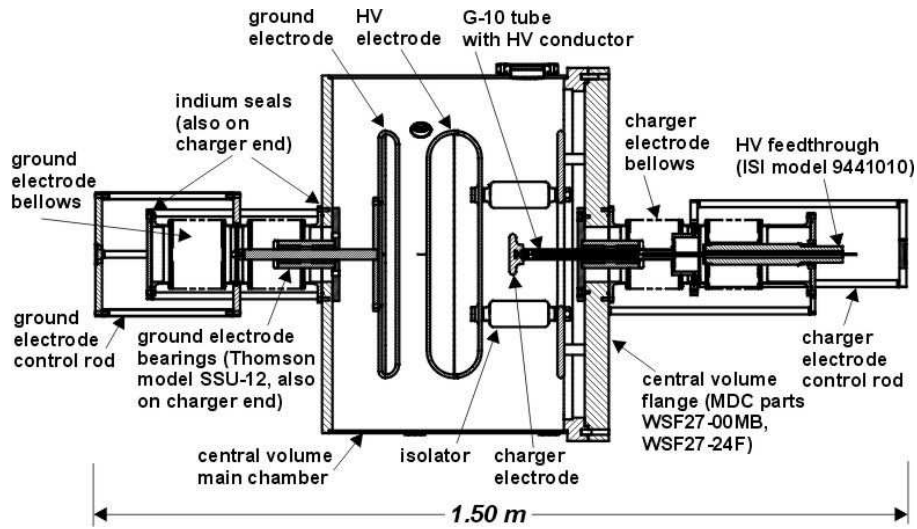}
\caption{\label{fig:centvol} Assembly drawing (elevation, to scale) of
  HV test system central volume.}
\vspace{1cm}
\end{center}
\end{figure}

The design chosen for the central volume of the test apparatus is shown
in Fig.~\ref{fig:centvol}.  It consists of a parallel-plate capacitor
with cylindrical electrodes, mounted horizontally on the
central axis of a cylindrical LHe tank.  

\begin{figure}[htbp]
\begin{center}
\includegraphics[width=12cm]{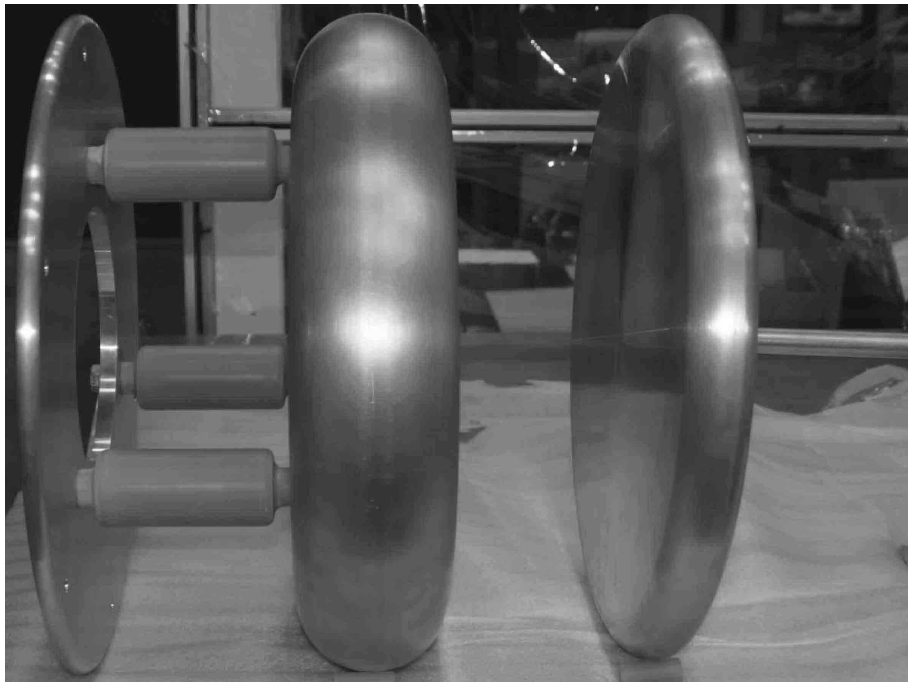}
\caption{\label{fig:electrodes} HV test system electrodes before
  installation in the central volume.  The HV electrode is in the
  center.  It is fixed to the solid aluminum plate at the left with three G-10
  insulating stand--offs.  The plate is in turn bolted (and grounded)
  to the inner surface of the central volume flange.}
\end{center}
\end{figure}

The electrodes of the parallel--plate
capacitor are shown in Fig.~\ref{fig:electrodes} and their dimensions
listed in Table~\ref{tab:electrodes}.  Both electrodes
consist of hollow cylindrical shells of type 6061 aluminum.  
Two small holes are bored into the outer edges of each electrode to 
allow for filling with LHe during operation.  

The HV electrode is fixed to the inner surface of the central volume end
flange with three insulating stand-offs.  The movable ground electrode
has the same diameter as the HV electrode but is narrower in profile.  
An aluminum rod attaches to the rear of the electrode and connects to
a welded bellows for position control along the axis normal to the
electrode surfaces.
\begin{table}
\caption{\label{tab:electrodes}HV test system electrode dimensions}
\begin{tabular}{|lr|}
\hline
Surface roughness, all electrodes&1.6~$\mu$m rms\\
Surface flatness, all electrodes&0.005~cm\\
Diameter, HV and ground electrodes&45.8~cm\\
Diameter, charger electrode&7.6~cm\\
Wall thickness, HV and ground electrodes&0.6~cm\\
Length, HV electrode&10.2~cm\\
Length, ground electrode&2.5~cm\\
Length, charger electrode&1.3~cm\\
Edge curvature radius, HV electrode&5.1~cm\\
Edge curvature radius (average), ground electrode&1.9~cm\\
Edge curvature radius, charger electrode&0.64~cm\\
Length, HV electrode insulators&14.0~cm\\
Diameter, HV electrode insulators&5.1~cm\\
\hline
\end{tabular}
\vspace{1cm}
\end{table}

The fixed HV electrode is charged via a second movable electrode
mounted on the side opposite the ground electrode.  It consists of a
solid type 6061 aluminum disk with rounded edges.  The electrode is 
attached to a G-10 tube which connects to a second bellows 
for position control.  A steel spring attaches to the back of the
electrode and leads out of the HV system through the G-10 tube to an
external power supply. 

The central LHe volume (Table~\ref{tab:central_volume}) of the test
system consists of a stainless steel vessel closed with a wire-seal
flange.  The flange is a standard part except that the number of bolt 
holes has been doubled to help insure against superleaks, and is itself a 
prototype component to be tested for the final version of the EDM
experiment.  To minimize heat loads, the central volume is held in
place with a sling made from two Kevlar ropes anchored in the
surrounding copper radiation shield
(Sec.~\ref{sec:vacuum_system}) with specially--designed jigs.
\begin{table}
\caption{\label{tab:central_volume} HV test system central LHe volume 
dimensions}
\begin{tabular}{|lr|}
\hline
Inner length (excluding flange spaces)&44.0~cm\\
Inner diameter (excluding flange spaces)&66.0~cm\\
Side wall thickness&0.5~cm\\
Welded end wall thickness&0.6~cm\\
Wire-seal flange thickness&4.4~cm\\
Bellows length (HV side)&37.3~cm\\
Bellows mean inner diameter (HV side)&10.8~cm\\
Mass&$\approx$300~kg\\
LHe volume (including flange spaces and bellows)&180~l\\
\hline
\end{tabular}
\vspace{1cm}
\end{table}

\begin{figure}[htbp]
\begin{center}
\includegraphics[width=12cm]{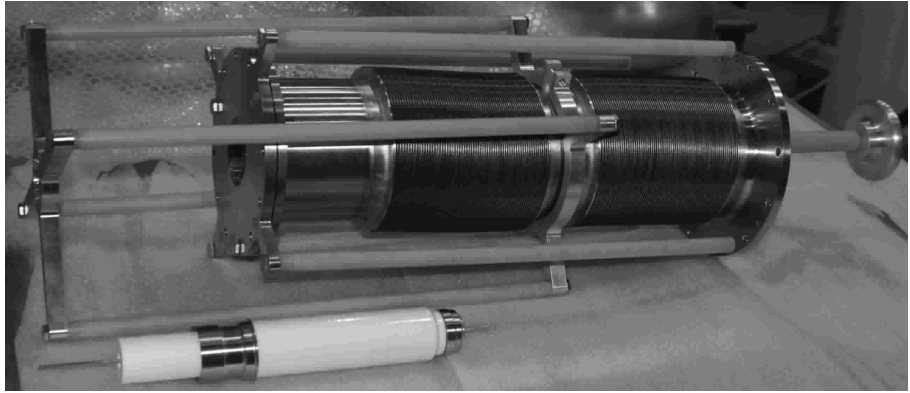}
\caption{\label{fig:bellows} Bellows system for position control of
  charger electrode, before attachment to HV test system central
  volume.  The charger electrode protrudes from the right end.  The ceramic
  feedthrough used to bring the initial HV potential into the LHe volume can
  be seen in the foreground before being welded into place in the small hole
  in the left end-cap of the bellows assembly.}
\end{center}
\end{figure}

Control of the movable ground and charger electrodes is provided by
welded bellows attached to either end of the central volume with indium seals.
Each bellows consists of two welded sections attached in
series with a rigid center ring between the sections
(Fig.~\ref{fig:bellows}). Rods attached to the movable electrodes exit
the central volume via bearings mounted in G-10 blocks, and connect to
the inner surfaces of
the center rings. During operation, the bellows fill with LHe through
a series of holes in the G-10 blocks.  The bearings are standard parts
with stainless steel balls in plastic casings from which all oil is
removed before installation.  Friction is a concern and operation of
the apparatus tests whether the LHe itself can serve as an
adequate lubricant. The bellows are operated by
control rods attached to the exteriors of the center rings, so that
one bellows section extends while the other compresses, maintaining a
constant LHe volume in the bellows.

\subsection{Vacuum System}
\label{sec:vacuum_system} 
\begin{figure}[htbp]
\begin{center}
\includegraphics[width=12cm]{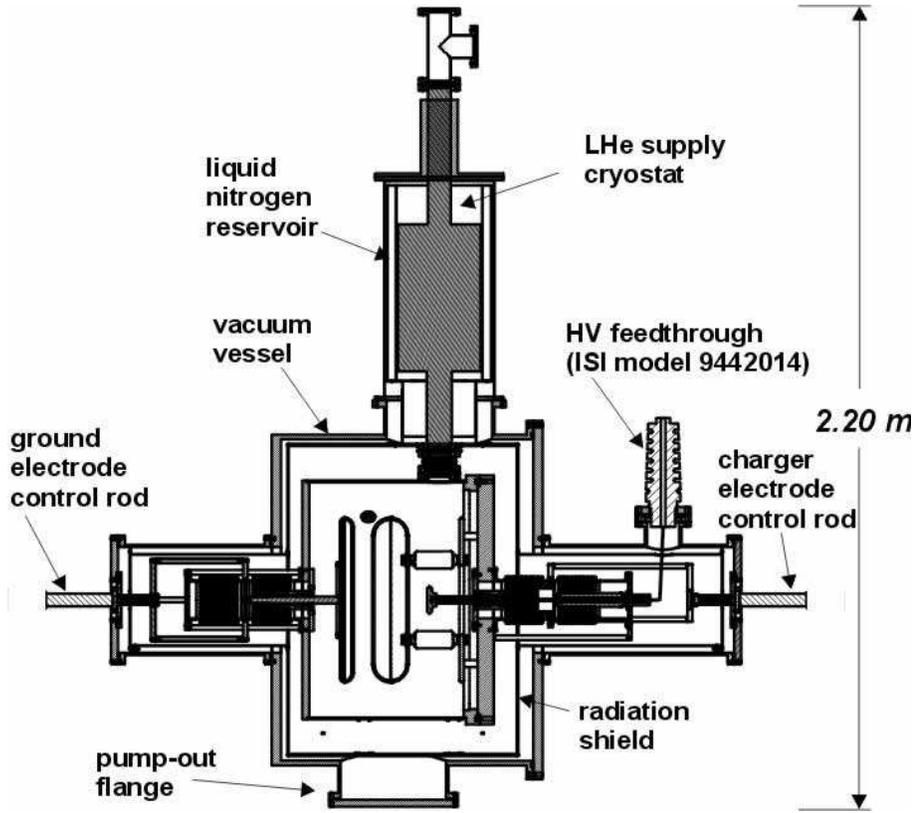}
\caption{\label{fig:vacsys} Assembly drawing (to scale) of HV test system
  vacuum volume.  Rods for electrode motion control exit the ends of
  the vacuum system via external bellows and connect to automated linear slides
  (Fig.~\ref{fig:system_photo}).}
\end{center}
\end{figure}

The complete design for the EDM HV test system is shown in
Fig.~\ref{fig:vacsys}.  The central volume is supplied with LHe via an 
auxiliary 20-liter cryostat mounted to the top.  The central and upper
LHe volumes are surrounded by thin-walled copper radiation shields,
cooled to about 100~K via thermal contact with a 10-liter liquid
nitrogen reservoir surrounding the upper volume.  All LHe volumes and
radiation shields are wrapped with 15--20 layers of aluminized Mylar
superinsulation.

The entire assembly is placed inside a $\sim$1~m$^{3}$ aluminum vacuum
chamber.  The chamber can typically be maintained at $1 \times 10^{-6}$
torr at room temperature with a $\sim$500~l~s$^{-1}$ turbomolecular pump
mounted at the bottom.  The vacuum has been observed to improve by more
than an order of magnitude with cryopumping.

\begin{figure}[htbp]
\begin{center}
\includegraphics[width=12cm]{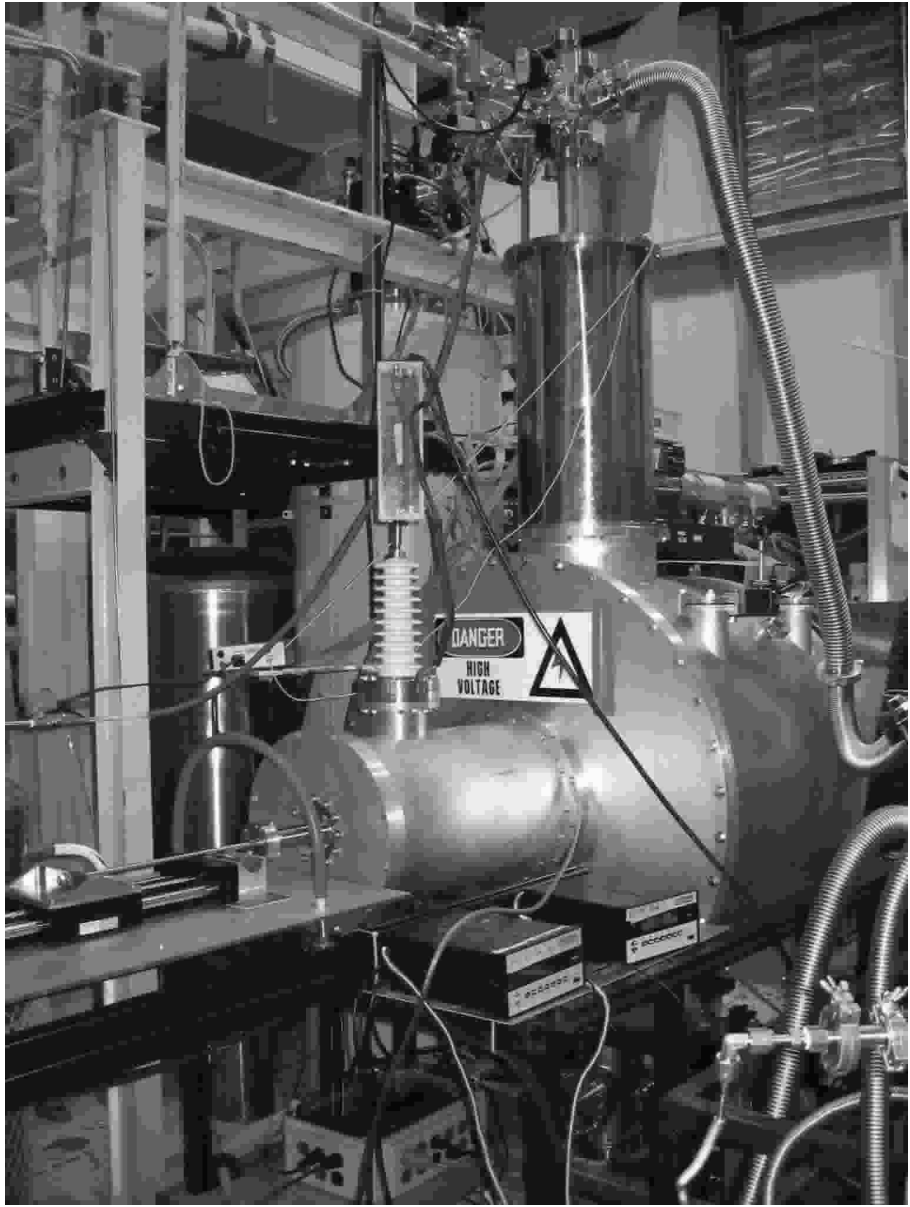}
\caption{\label{fig:system_photo} Fully assembled HV test
  system. The vacuum system arm surrounding charger electrode control
  bellows is
  visible in the foreground, with the air-vacuum HV feedthrough attached at
  the top.  The small external bellows, control rod, and motor--driven linear
  slide for the charger electrode control are visible in front of the
  arm.  The supply
  cryostat is visible on top of the main vacuum volume.  The flexible tubing
  for pumping on the LHe bath stretches down from the top of the
  supply cryostat.}
\end{center}
\end{figure}

The HV interface between air and LHe consists of two standard ceramic
feedthroughs (Table~\ref{tab:feedthrough}).  One feedthrough is mounted to a
flange on the outer arm of the vacuum system above the charger 
electrode control bellows 
(Figs.~\ref{fig:vacsys} and ~\ref{fig:system_photo}), and a smaller
feedthrough is welded into the end-cap of the HV bellows
(Fig.~\ref{fig:bellows}). The upper ceramic portion of the small
feedthrough protrudes completely into the LHe volume
(Fig.~\ref{fig:centvol}); operation of the system tests its
performance in the cryogenic environment.  The bottom tips of
the feedthrough conductors are soldered together, and the
upper tip of the smaller feedthrough is soldered to
a stainless steel spring.  The spring attaches to the charger
electrode through the G-10 tube.
\begin{table}
\caption{\label{tab:feedthrough}HV feedthrough parameters}
\begin{tabular}{|lr|}
\hline
Air--vacuum feedthrough ceramic length&29.2~cm\\
Air--vacuum feedthrough ceramic outer diameter&8.9~cm\\
Air--vacuum feedthrough breakdown rating&100~kV\\
Vacuum--LHe feedthrough ceramic length&24.0~cm\\
Vacuum--LHe feedthrough ceramic outer diameter&3.8~cm\\
Vacuum--LHe feedthrough breakdown rating&40~kV\\
\hline
\end{tabular}
\vspace{1cm}
\end{table}

The control rods for the ground and charger electrodes attach to the
inner end-caps of small welded bellows mounted externally to either end
of the vacuum system (Fig~\ref{fig:system_photo}).  The external bellows are
in turn attached to screw--driven linear slides controlled by servo
motors.  With this system, the minimum gap between the HV and
ground (charger) electrodes can be adjusted from 0 to 10~cm (7~cm) over a wide
range of speeds, with a nominal accuracy of less than 0.1~mm.  The gap
between the electrode surfaces is viewable through a series of 5~cm diameter
quartz view ports in the side walls of the central volume and vacuum
chamber.  

\subsection{Data Acquisition}
\begin{figure}[htbp]
\begin{center}
\includegraphics[width=12cm]{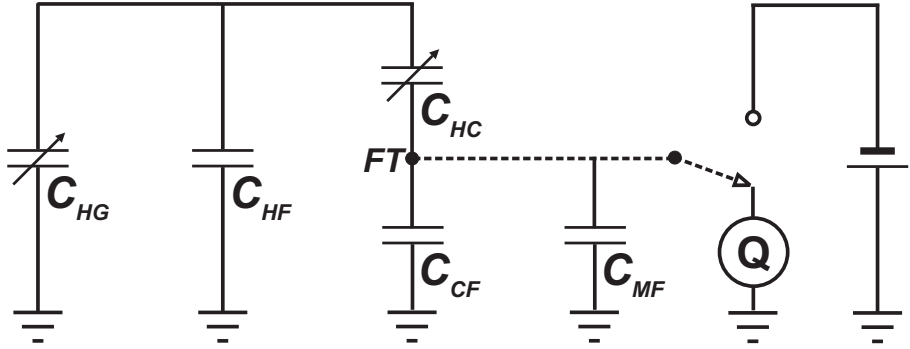}
\caption{\label{fig:circuit} HV test system equivalent circuit. The
  point labeled {\it FT} corresponds to the upper tip of the 
  air-vacuum feedthrough and forms the output of a capacitive
  divider.  One side of the divider is formed by the capacitance
  between the HV and ground electrode ($C_{HG}$), in series with the
  capacitance of the HV and charger electrode ($C_{HC}$).  The other
  side is formed by the fixed capacitance of the charger electrode and 
  ground ($C_{CF}$).  $C_{HF}$ is the fixed capacitance of the HV
  electrode with respect to ground.  When the system is connected to a
  charge--sensitive meter (Q), $C_{CF}$ is in parallel with the fixed
  capacitance of the cable plus the meter with respect to ground ($C_{MF}$)}  
\end{center}
\vspace{1cm}
\end{figure}

The view ports are designed to permit the eventual use of a
laser to monitor the electric field via the Kerr effect in the LHe
\cite{Sushkov04}.  For the present studies, the small charger electrode
is used as a
capacitive probe of the voltage between the HV and movable ground 
electrodes.  The equivalent circuit of the HV apparatus is shown in
Fig.~\ref{fig:circuit}.  In a typical HV amplification and
measurement cycle, a 50~kV HV power supply is connected to the air--vacuum 
feedthrough via a 6~m cable.  The gap
between the ground and HV electrodes is set to an initial value of a
few millimeters.  The charger electrode is brought into contact with
the HV electrode, and a potential of a few tens of kV is applied.
With the power supply still engaged, the charger is retracted to a
reference position several centimeters behind the HV electrode.  The
power supply is switched off, and the cable is moved from the power
supply output to the input of a charge-sensitive meter.  The meter records
the change in charge $\Delta Q_{C}$ on the charger electrode as the
gap between the HV and ground electrode is increased to amplify the voltage. 

From the diagram in Fig.~\ref{fig:circuit}, the change in voltage
across the variable capacitance $C_{HG}$ is given by:
\begin{equation}
\Delta V_{HG} = \Delta Q_{C}\left(\frac{1}{C_{HC}} + \frac{1}{C_{CF} +
    C_{MF}}\right)
\label{eq:potential}
\end{equation}
In practice, a sensitive current amplifier is used in place of a
charge meter and the output is integrated over time during the measurement. 

\begin{figure}[htbp]
\begin{center}
\includegraphics[width=12cm]{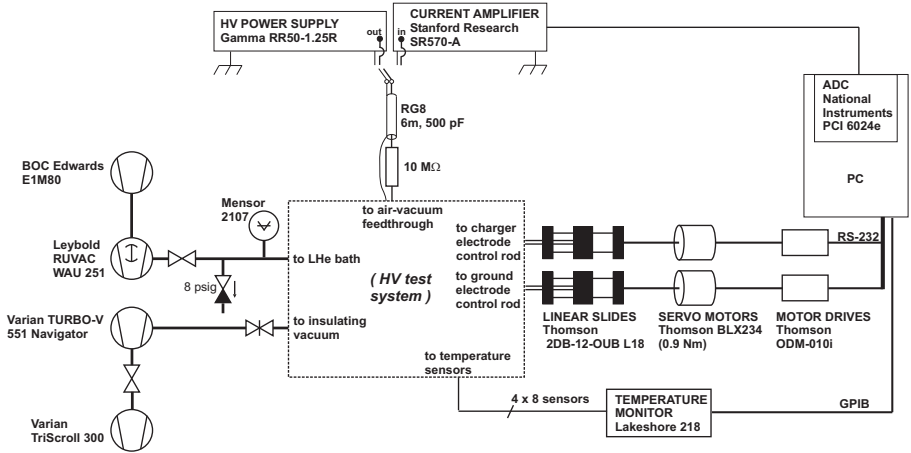}
\caption{\label{fig:daq} Basic schematic of complete system.}
\end{center}
\end{figure}

The complete schematic of the HV test system is shown in Fig.~\ref{fig:daq}.
A 10~M$\Omega$ resistor is put in series with the cable to provide
some protection for the electronics from spark discharges.  For data
acquisition after charging, the cable is switched manually to the
current amplifier.  The current amplifier is read out with a PC-based
digital--to--analog converter, at a software-limited rate 
(for this work) of 130 Hz.  For diagnostic purposes during cool-down, 
temperatures are monitored at various points on the central volume and 
radiation shield with a
series of silicon diode sensors.  During operation, LHe temperature
is controlled by pumping on the main bath with a 25~l~s$^{-1}$ (air) 
oil-sealed mechanical pump (occasionally boosted to $\sim$
40~l~s$^{-1}$ with a roots
blower), and temperature is inferred from the He 
vapor pressure.  The motors running the linear slides are
PC-controlled via servo drives.  Initial HV--ground electrode spacing
is measured with a surveyor's transit focused on the electrode
surfaces through the view ports.

\section{Results}
\subsection{Cryogenic Performance and HV Test Conditions}
Conduction cool-down of the radiation shields from room temperature to
100~K takes approximately 72~hr.  Liquid nitrogen pre--cooling of the
central volume is done in steps over the course of about 12~hr so as
to minimize thermal gradients across the large wire--seal flange.
This is further aided by spraying the liquid in radial jets onto the 
side walls of the central volume and relying on the
thermal conductivity of the walls to cool the flange uniformly.
Initial filling of the LHe main volume from a standard 500~l storage
dewar takes approximately 45~min.  SF production is achieved by
pumping on the main bath.  To ensure quantities of SF sufficient to
completely immerse the electrodes are obtained,  both the LHe supply
and HV system volumes are pre--cooled via pumping to just above the
lambda point.  The HV system bath is then topped off with LHe at low
vapor pressure ($\sim 90$~torr) before being pumped to below the lambda point. 

During normal operation, typical LHe boil--off rates of 2--3~l~hr$^{-1}$
(liquid) are observed.  This rate corresponds to a total heat load of
1.6- 2.3~W,
likely dominated by conduction through the upper neck of the small supply
cryostat.  No appreciable changes in the insulating vacuum are observed
(resolution $\sim 10^{-7}$~torr) during SF production or operation,
indicating the feasibility of using the standard wire--seal flange
technology for large volumes of SF in the EDM cryostat.  No
degradation in the range and precision of the position control of the
electrodes is observed at cryogenic temperatures.  However, moving the
electrodes requires significantly higher ($\sim$ factor 3) torque from the
motors than at room temperature.  While the reason is unclear, this
factor remains constant over the course of operation of the system,
suggesting a stiffening of the bellows as opposed to an increase in
bearing friction.

Tests are performed on saturated LHe with vapor pressures ranging
from 27.3 to 880 torr, corresponding to temperatures of 2.05~K to
4.38~K.  By the two--fluid model, SF concentrations reach as
high as 25\% at the lowest temperature attained.  The lambda transition
consistently occurs at 36 torr (2.16 K) in the system, inferred from
the rapid and complete cessation of turbulence and rising He vapor
bubbles observable in the view ports.

\subsection{Calibration}
Use of Eq.~\ref{eq:potential} requires knowledge of the HV--charger
capacitance ($C_{HC}$) and fixed capacitance ($C_{CF}$) when the gap 
between these electrodes is set to
its reference value at which the electrical measurements are made.  
The reference spacing is 5.0~cm, as given by the motor driver
software.  With the system full of LHe just below the lambda point,
the HV electrode is grounded by closing the gap between it and the movable 
ground electrode.  A digital capacitance bridge is connected between
the upper tip of the air--vacuum feedthrough and the
nearest point on the exterior of the (grounded) insulating vacuum
chamber.  Bridge readings are then recorded as the charger is moved
back from the HV electrode.

\begin{figure}[htbp]
\begin{center}
\includegraphics[width=12cm]{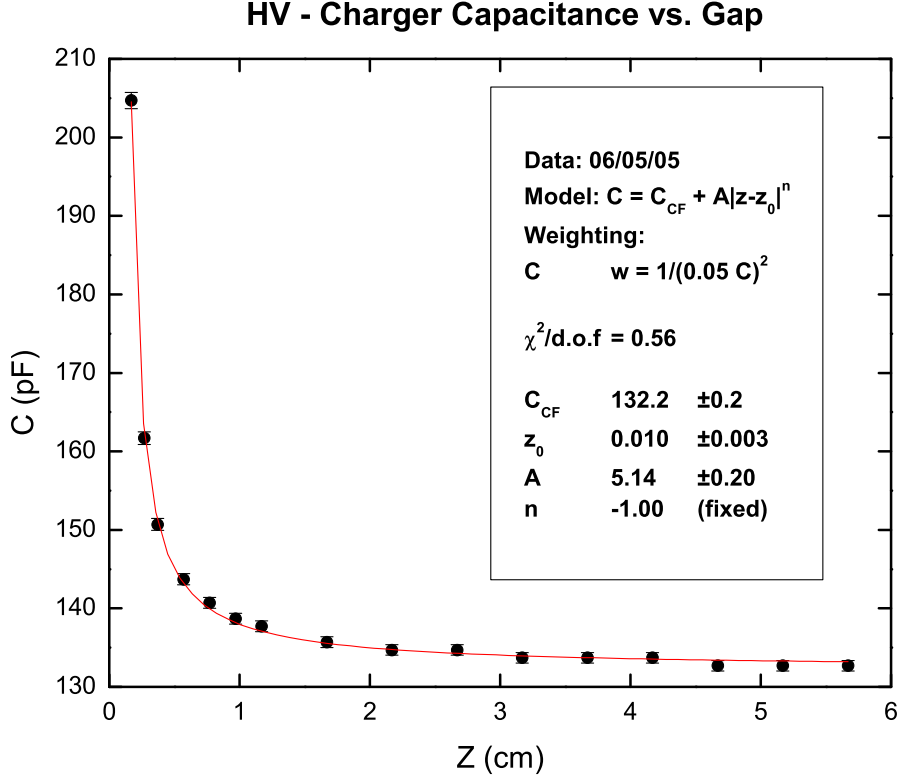}
\caption{\label{fig:chc} Charger--HV electrode capacitance as a
  function of separation.  A statistical error of 0.5\% is assumed on each
  point.  The legend shows the result of a least-squares fit.}
\end{center}
\end{figure}

Results are shown in Fig.~\ref{fig:chc}.  Data represent the parallel sum of
the capacitances $C_{HC}$ and $C_{CF}$ (Fig~\ref{fig:circuit}).  To
extract separate values for these capacitances, $C_{HC}$ is modeled as
a parallel-plate capacitor and the data are fit the expression
\begin{equation}
C = C_{CF} + C_{HC} = C_{CF} + \frac{A}{z-z_{0}}.
\end{equation}
Here, $A$ is a constant, $z$ is
the electrode separation, and $z_{0}$ is an offset reflecting the
uncertainty in the absolute separation.  The fit returns a value of
$132.2\pm0.2$~pF for $C_{CF}$.  Using the values of $A$ 
and $z_{0}$ returned from the fit and substituting $z = 5.0$~cm for
the reference separation yields the reference value for 
$C_{HC}$ of 1.05$\pm$.05~pF.  As $C_{HC}$ is less than 1\% of $C_{CF}$
(and in any case $C_{HC}<<C_{MF}$), Eq.~\ref{eq:potential} reduces to
\begin{equation}
\Delta V_{HG} = \left(\frac{1}{C_{HC}}\right)\int i dt
\label{eq:potential_used}
\end{equation}
where $i$ is the output of the current amplifier.

\begin{figure}[htbp]
\begin{center}
\includegraphics[width=12cm]{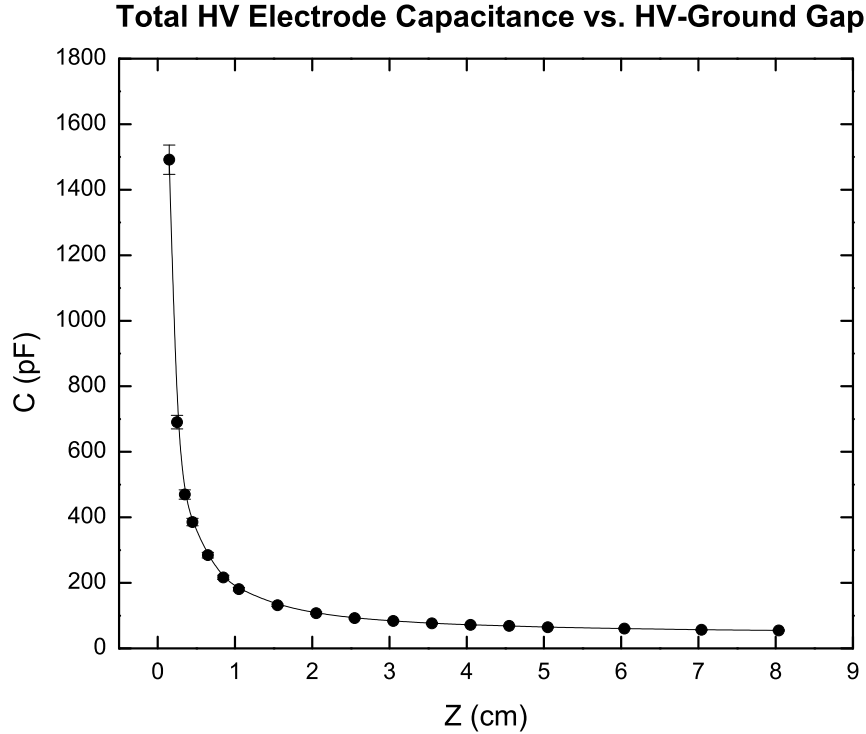}
\caption{\label{fig:chv} Total HV electrode capacitance as a
  function of HV-ground electrode separation.  A statistical error of 3\%
  is assumed on each point.  The line connecting the points is a
  spline interpolation.}
\end{center}
\end{figure}

An assessment of the leakage current from the HV electrode at a given
HV--ground electrode gap (Sec.~\ref{sec:ileak}) requires a knowledge of the
total capacitance of the HV electrode with respect to ground
at the particular gap setting.  This capacitance is very closely
approximated by $C_{HG} + C_{HF}$ in Fig.~\ref{fig:circuit}.
With the bridge still connected as
above, both the ground and charger electrodes are brought into contact
with the HV
electrode.  Bridge readings are then recorded as the ground electrode
is stepped
back from the HV electrode.  The data are shown in Fig.~\ref{fig:chv}.
This curve yields a direct measurement of the desired capacitance as a
function of the gap.

\subsection{Voltage Amplification and Breakdown Strength}
Amplification tests are performed with an initial gap between
the HV and ground electrodes of 2--3~mm.  At this gap, the highest
possible potential is
established with the power supply.  This is typically achieved after
several attempts in which
spark discharges occur across the
gap at lower potentials.  Presumably these discharges are initiated by 
sharp points and/or frozen contaminants on the electrode surfaces, and they
decrease in frequency as contaminants are destroyed by 
successive sparks in a process analogous to the ``conditioning''
observed in vacuum--insulated systems.  During tests with the
initial gap set to a few cm, the HV electrode can usually be 
charged to the full capacity of the power supply (50~kV) with the
system full of LHe above or below the lambda point.  This indicates
that the vacuum-LHe feedthrough can withstand voltages
25\% higher than its nominal breakdown rating in air when immersed in LHe.

\begin{figure}[htbp]
\begin{tabular}{p{7cm}p{7cm}}
\hspace{0cm} {\large (a)} & \hspace{0.5cm} {\large (b)} \\
\includegraphics[width=6.6cm]{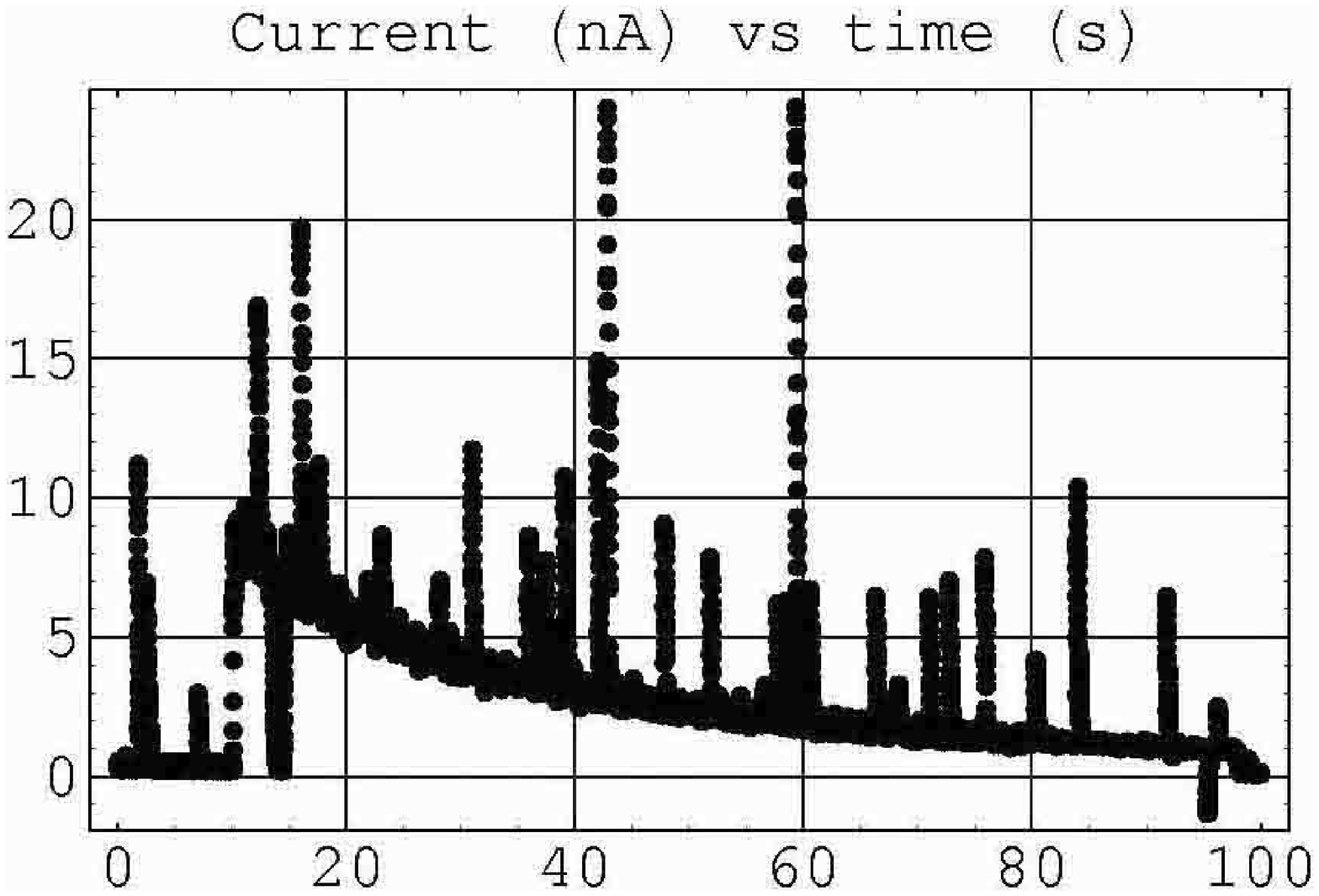} &
\includegraphics[width=6.6cm]{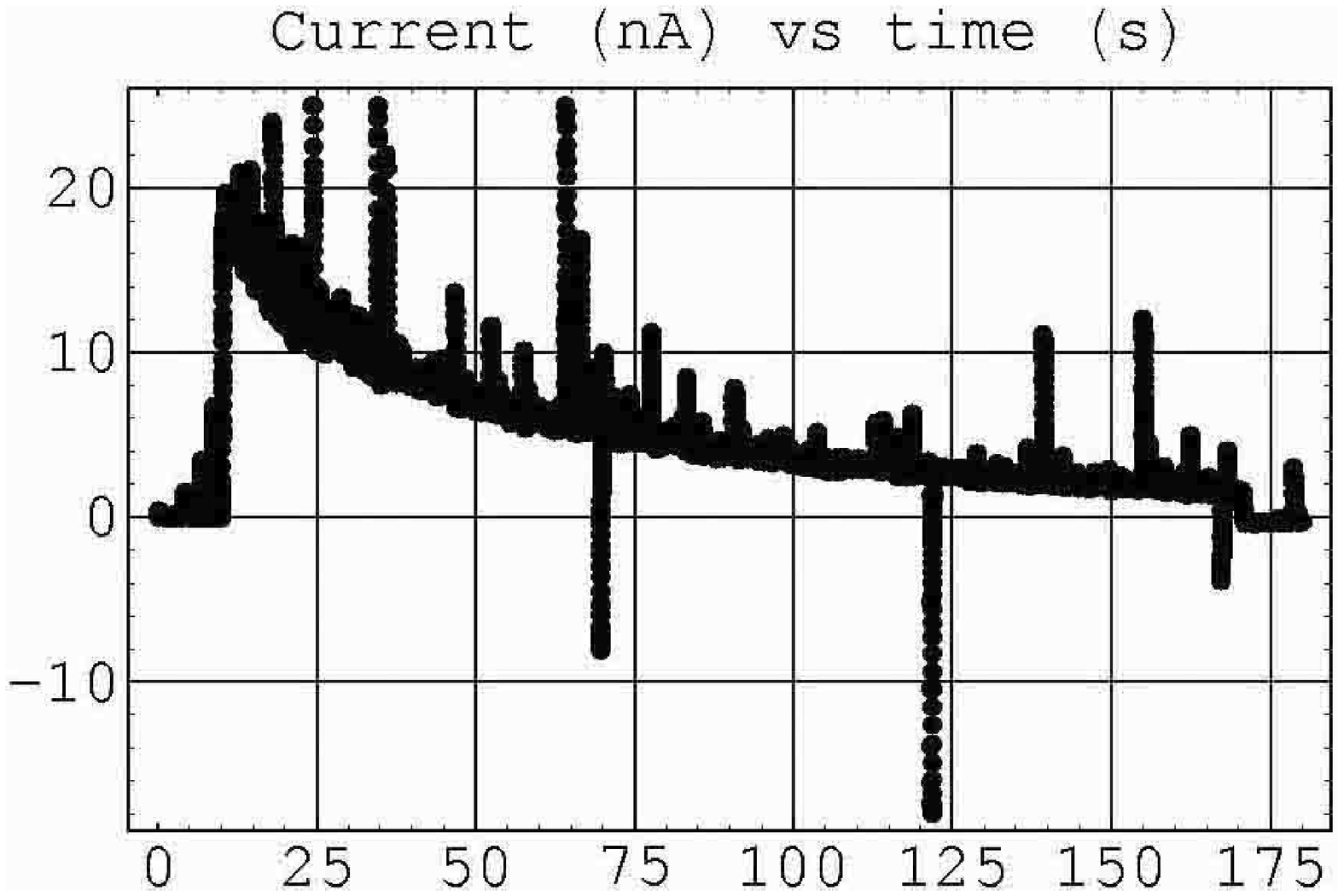}\\
\hspace{0cm} {\large (c)} & \hspace{0.5cm} {\large (d)} \\
\includegraphics[width=6.6cm]{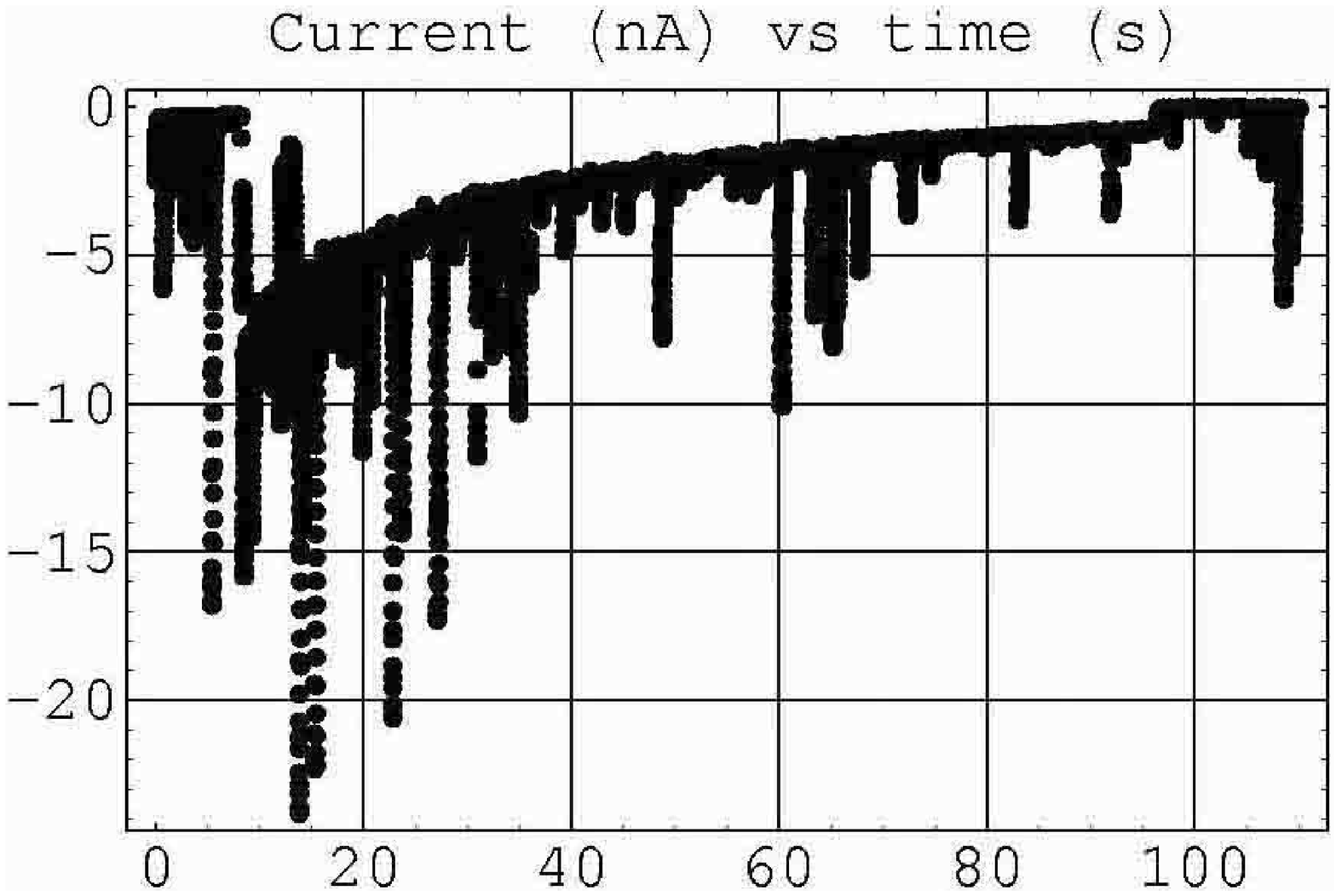} &
\includegraphics[width=6.6cm]{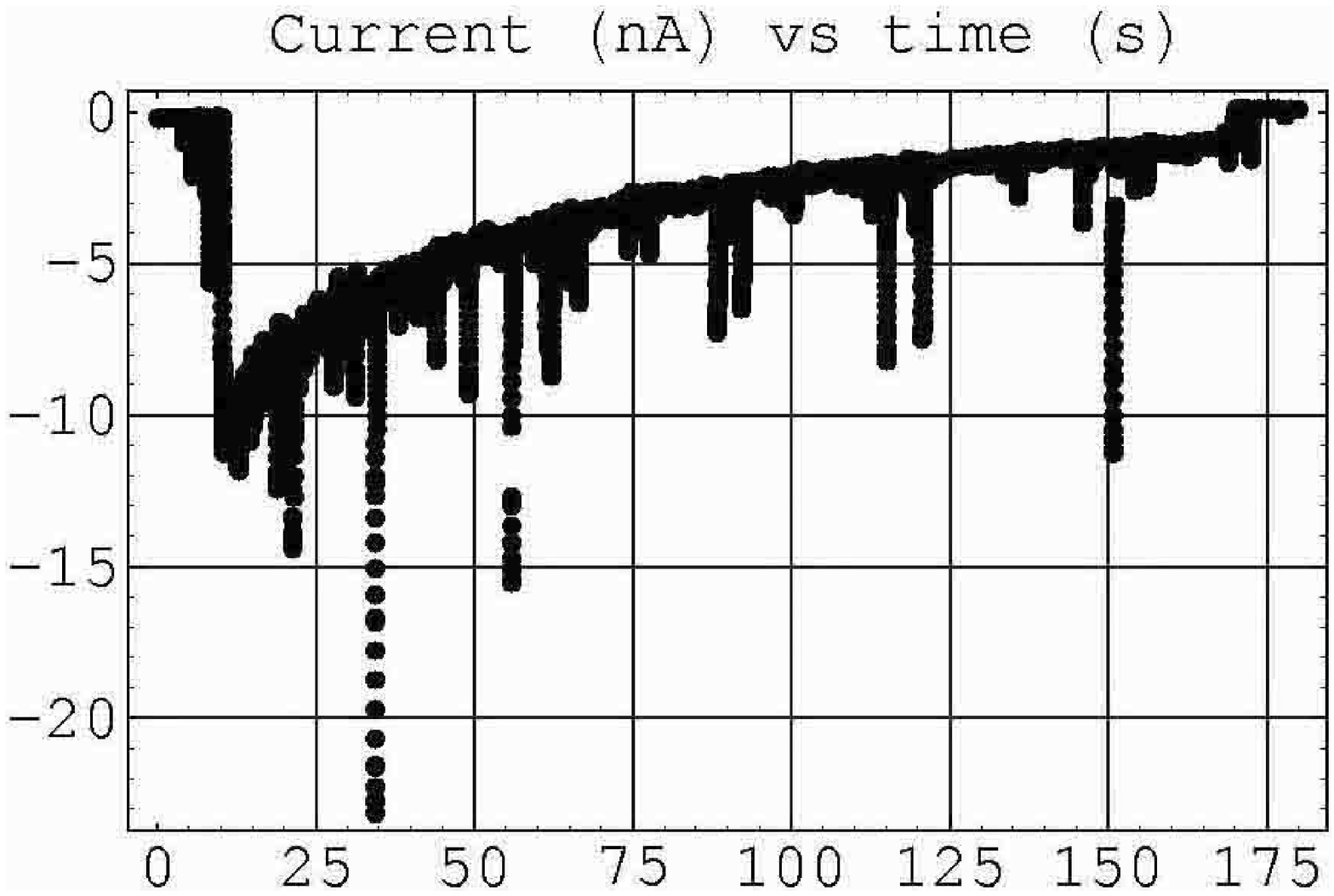}\\
\end{tabular}
\caption{\label{fig:raw_data} Amplification data showing output of
  the current amplifier in nA vs. time in seconds.  Data sets
  correspond to the largest voltages attained with the test system.
  In (a), the data are obtained with the LHe bath at 34.5 torr (2.14~K), an
  initial gap (visible in the transit) of 2.9~mm, and an initial
  potential of 13~kV at this gap.  The ground electrode is retracted
  at 5.08~cm~min$^{-1}$. In (b), the LHe bath pressure is 880 torr (T =
  4.38~K), the initial gap is 3.0~mm, the initial potential is
  42~kV, and the electrode retracted at 2.54~cm~min$^{-1}$. In (c),
  the LHe bath pressure is 
  31.9 torr (T = 2.11~K), the initial gap is 3.0~mm, the initial potential is
  -11.5~kV, and the electrode retracted at 5.08~cm~min$^{-1}$. In (d),
  the LHe bath
  pressure is 654 torr (T = 4.06~K), the initial gap is 3.1~mm, the
  initial potential is
  -31~kV, and the electrode retracted at 2.54~cm~min$^{-1}$.}
\end{figure}

Voltage amplification data are shown
in Fig~\ref{fig:raw_data}.  Output of the current amplifier is plotted
as a function of the time since the start of a data acquisition cycle.
With the exception of a few transients, the current traces in 
Fig~\ref{fig:raw_data} remain at zero for the first few seconds after 
the data acquisition is started, consistent with no current
flow---that is, no transfer of charge to the charger electrode---when
the ground electrode is at rest.  When the ground electrode 
retraction is begun (after approximately 10~s) the current immediately 
rises (for example, to about 10~nA in the case of Fig~\ref{fig:raw_data}a).
The current then slowly decays (for example, over the next 85~s in 
Fig~\ref{fig:raw_data}a) as the retraction proceeds at constant
velocity.  In the limit of an ideal parallel--plate capacitor, the
current would be expected to remain constant for a constantly
increasing plate separation; the decay observed in
Fig~\ref{fig:raw_data} is a consequence of the fixed capacitance
between the HV electrode and ground ($C_{HF} \sim 40$~pF, as inferred
from bridge measurements).  The current drops
back to zero when the ground electrode retraction stops (for example,
after 95~s, corresponding to a gap of 7.8~cm, in Fig~\ref{fig:raw_data}a). 

\begin{figure}[htbp]
\begin{center}
\includegraphics[width=12cm]{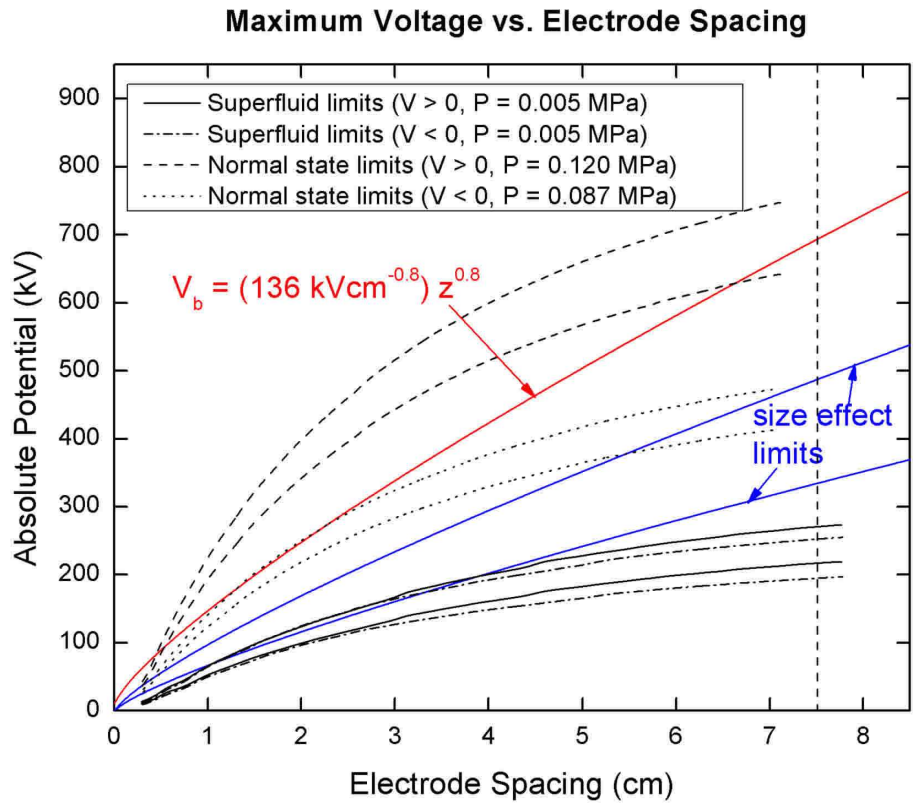}
\caption{\label{fig:hvdata} Voltages obtained in HV test
  system as function of electrode spacing.  The data curves are
  labeled only in the legend.  The data curves define the
  range (as determined by the 1-sigma error bars) of the maximum
  voltages obtained in LHe, for positive and negative initial
  potentials,
  above and below the lambda point. The extrapolated
  breakdown curve and size effect breakdown estimates from
  Fig.~\ref{fig:projections} are also shown.}
\end{center}
\vspace{1cm}
\end{figure}

The total charge $\Delta Q_{C}$ accumulated on the charger
electrode, as the ground electrode is retracted, is computed from the sum
\begin{equation}
\Delta Q_{C}=\sum_{n}(t_{n+1}-t_{n})(i_{n+1}+i_{n})/2,
\label{eq:charge}
\end{equation}
where $t_{n}$ and $i_{n}$ represent the time and current values of the
$n$th point in the data set.  A value for Eq.~\ref{eq:potential_used} is 
obtained by dividing this sum by $C_{HC}$, and the total voltage is
found by adding the value of the initial potential between the
plates.  The voltage at any
value of the electrode separation is found by truncating the sum in
Eq.~\ref{eq:charge} at each time point and converting the time axis to
electrode separation.  The voltage as a function of separation derived
from the data in Fig.~\ref{fig:raw_data} is shown in
Fig.~\ref{fig:hvdata}.  In each case, the range of values defined by
the 1-sigma error bars is shown. 

The extrapolated breakdown and size effect curves from
Fig.~\ref{fig:projections} are also shown for reference.  Direct 
comparison between the data and the extrapolated breakdown
curves is misleading in that, first, the data curves
represent the highest voltages successfully maintained in the system
without breakdown during single, complete sweeps of the ground
electrode. The shape of the data curves is determined by
the capacitance of the high voltage electrode with respect to ground
as the spacing
changes, and not by breakdown events observed at each value of
the spacing.  As expected for the parallel plate geometry, the voltage
increase is approximately linear with the spacing for the first
1-2~cm, then levels off in the presence of the large fixed
capacitance.  Second, the
normal state data suggest that voltages larger than the
dielectric strength measured in previous experiments are attained below
1~cm (see Fig.~\ref{fig:projections}).  Here it is important
to note that the normal state data (positive voltage) 
in Fig.~\ref{fig:hvdata} are obtained at a pressure of 880 torr.  
This is approximately 20\% higher than for the data in
Fig.~\ref{fig:projections} and, as described in the next section,
dielectric breakdown in LHe is a strong function of pressure.  Normal
state data with negative voltage in Fig.~\ref{fig:hvdata} are
obtained just above atmospheric pressure in the lab at Los Alamos and
most closely approximate the conditions of the experiment in
Fig.~\ref{fig:projections}.

Three contributions to the error bars are considered.  First, a 5\%
uncertainty results from the error in $C_{HC}$, as derived from the
fit in Fig.~\ref{fig:chc}.  Second, the output from the current amplifier
corresponding to zero current is observed
to drift by as much as 0.1~nA over several tens of seconds, based on 
observations of the tails of the raw data plots 
(Fig.~\ref{fig:raw_data}) corresponding to when the ground electrode
is at rest.  An average zero offset of this magnitude 
translates into a $\pm$3\% shift in the summation 
in Eq.~\ref{eq:charge}, and a corresponding shift in the measured 
voltage.  The third (and usually the largest) effect is due to the transient
spikes observed in the raw data traces.  While it is unlikely that 
the transients represent an actual accumulation or loss of charge on the
electrodes during amplification and can be ignored for the purpose of
determining the voltage (see Sec.~\ref{sec:transients} below), the
evidence as of this
writing is anecdotal, therefore the effect of the transients is 
accounted for in the uncertainty.  To determine the effect of the
transients, the data sets in Fig.~\ref{fig:raw_data} are partitioned into
1~s time bins.  The minimum absolute current is determined in each
bin, and all individual points in a given bin with current values
greater than 20\% of the minimum are removed.  Eq.~\ref{eq:charge} is
used again to find the voltage vs. separation from the reduced data set,
which is then subtracted from the value obtained from the original
data set.  For the case of positive initial potentials, removal of 
transients from the SF (normal state) data reduces the
voltage by an average of 13\% (6\%).

The effect of the transients is positive (negative) in the case of
positive (negative) initial HV electrode potential, so the contribution of
the transients to the error bars is one--sided.  At maximum electrode
separation (7.8~cm for the measurement in SF and
7.2~cm for the measurement in normal state LHe), the largest positive
voltages attained are 240$^{+34}_{-14}$~kV and 688$^{+58}_{-41}$~kV,
respectively.  The largest negative voltages are
-215$^{-37}_{+13}$~kV and -443$^{-31}_{+27}$~kV, respectively. These
values represent lower limits on LHe breakdown
strength below and above the lambda point at large electrode separations; the
stressed LHe volumes are 12.8 liters for SF and 11.8 liters
for the normal state.

\subsection{Pressure Dependence of Breakdown Strength}
\begin{figure}[htbp]
\begin{center}
\includegraphics[width=12cm]{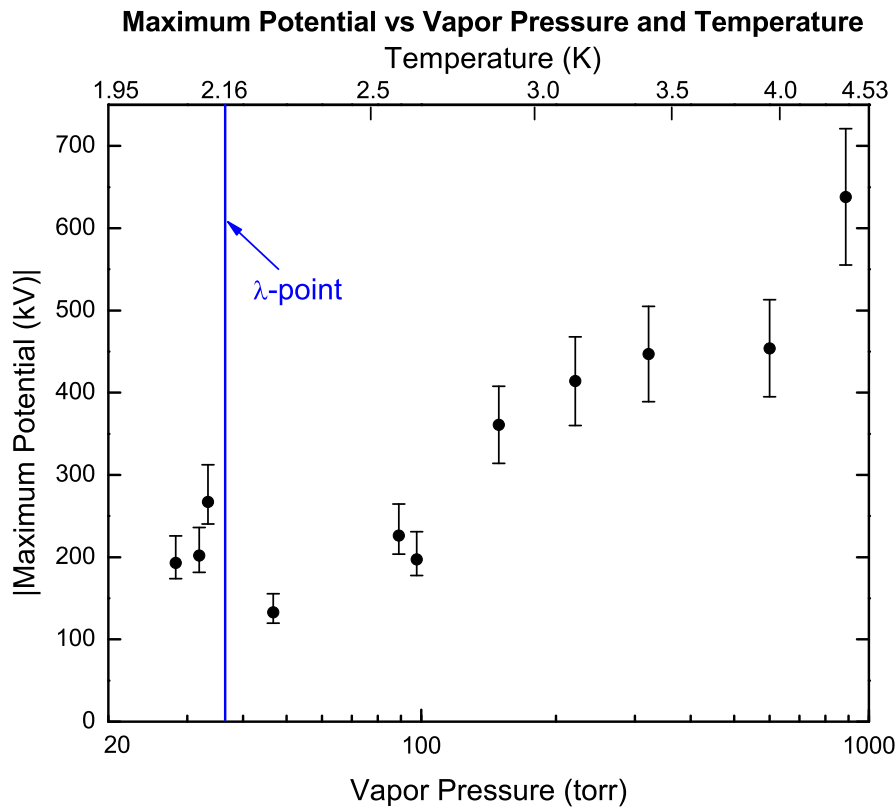}
\caption{\label{fig:VvsPT} Maximum voltages attained in HV test system
  over 12 hour period as a function of pressure and
  temperature.  Statistics below lambda point are highest values
  measured in 0.05~K bins above 27.3 torr
  (2.05~K) obtained in a series of many tests over a 6 hour period. Remaining
  data are obtained over subsequent 6 hours in which system is warmed
  continuously by the $\sim$~2~W heat load.} 
\end{center}
\end{figure}

The maximum voltages attainable in the HV test system below the lambda point
are about 30-50\% of those attained above, and are roughly 60\% of the design
goal for the EDM experiment.  It is unlikely that the reduced
performance of the HV test system below the lambda point is due exclusively to
an intrinsic property of the SF component of the LHe.  This is
illustrated in Fig.~\ref{fig:VvsPT}, which shows the maximum
voltages attained in the system at various pressures and temperatures. 

From the figure, comparable maximum voltages are attainable in the system
both above and below the lambda point at vapor pressures of 100 torr
and below.  This is consistent with several previous reports of
comparable breakdown strength of saturated normal state and
SF LHe \cite{Gerhold98_2}.  While the data above the lambda
point were taken over the course of a 6 hour warm up period, this
period occurred immediately after an interval of comparable
duration in which the system was kept below the lambda point and
during which many breakdown tests were performed.  Therefore it is
unlikely that the observed improvement with pressure is the result of
conditioning.  It seems reasonable that, as in at least one previous
experiment \cite{Gerhold98_2}, some of the resistance against
breakdown observed in the system near atmospheric pressures can be
recovered by pressurizing a sealed volume of LHe kept below the lambda
point.  Based on the compressibility of SF LHe
($10^{-7}$~Pa$^{-1}$ \cite{Keller69}), the pressurization would only require a
variable volume of a few liters in series with the HV test system main
volume and is a possible design improvement in the near future.

\subsection{Leakage Current Limits}
\label{sec:ileak}
\begin{figure}[htbp]
\begin{tabular}{p{7cm}p{7cm}}
\hspace{0cm} {\large (a)} & \hspace{0.5cm} {\large (b)} \\
\includegraphics[width=6.6cm]{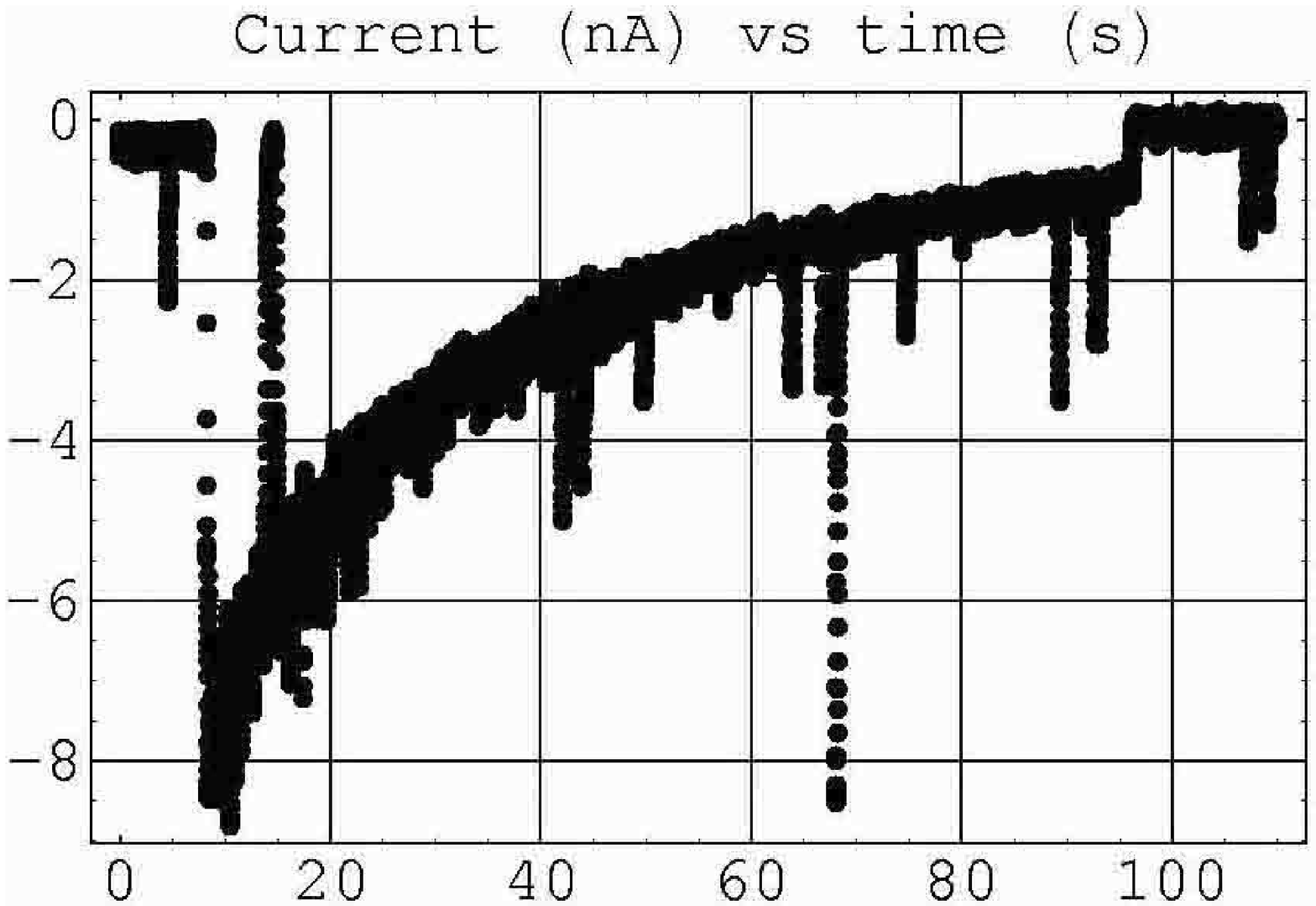} &
\includegraphics[width=6.6cm]{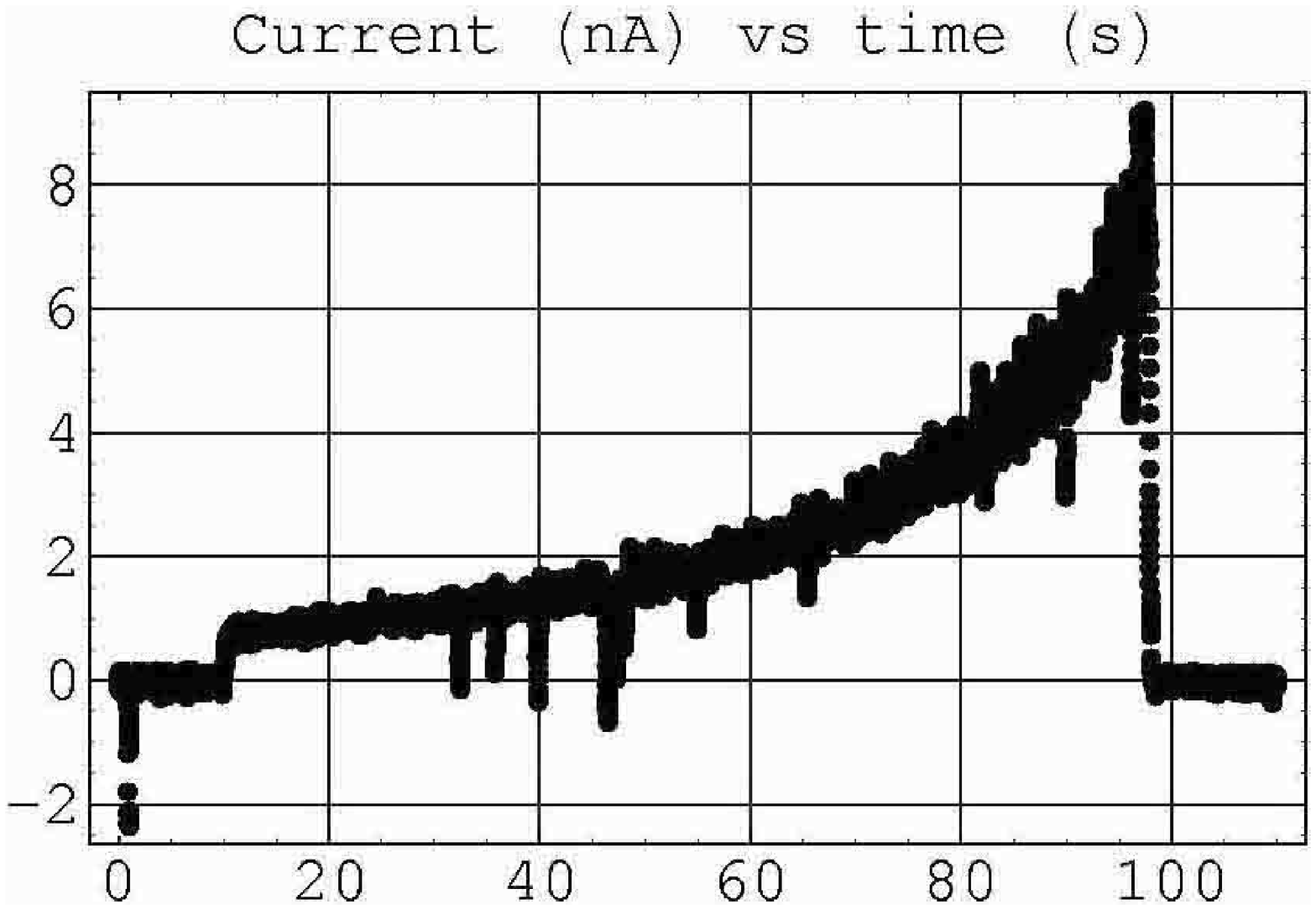}\\
\hspace{0cm} {\large (c)} & \hspace{0.5cm} {\large (d)} \\
\includegraphics[width=6.6cm]{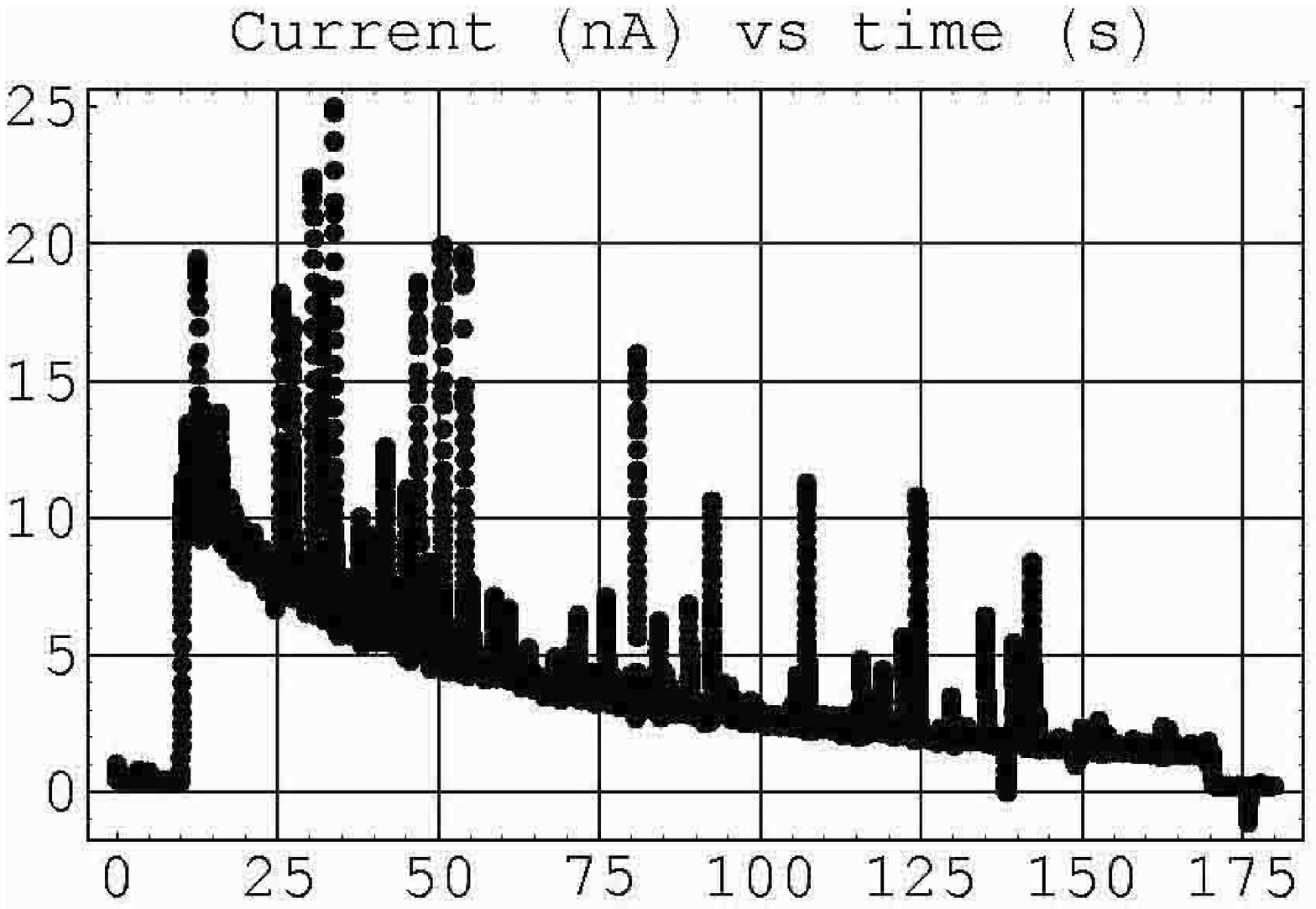} &
\includegraphics[width=6.6cm]{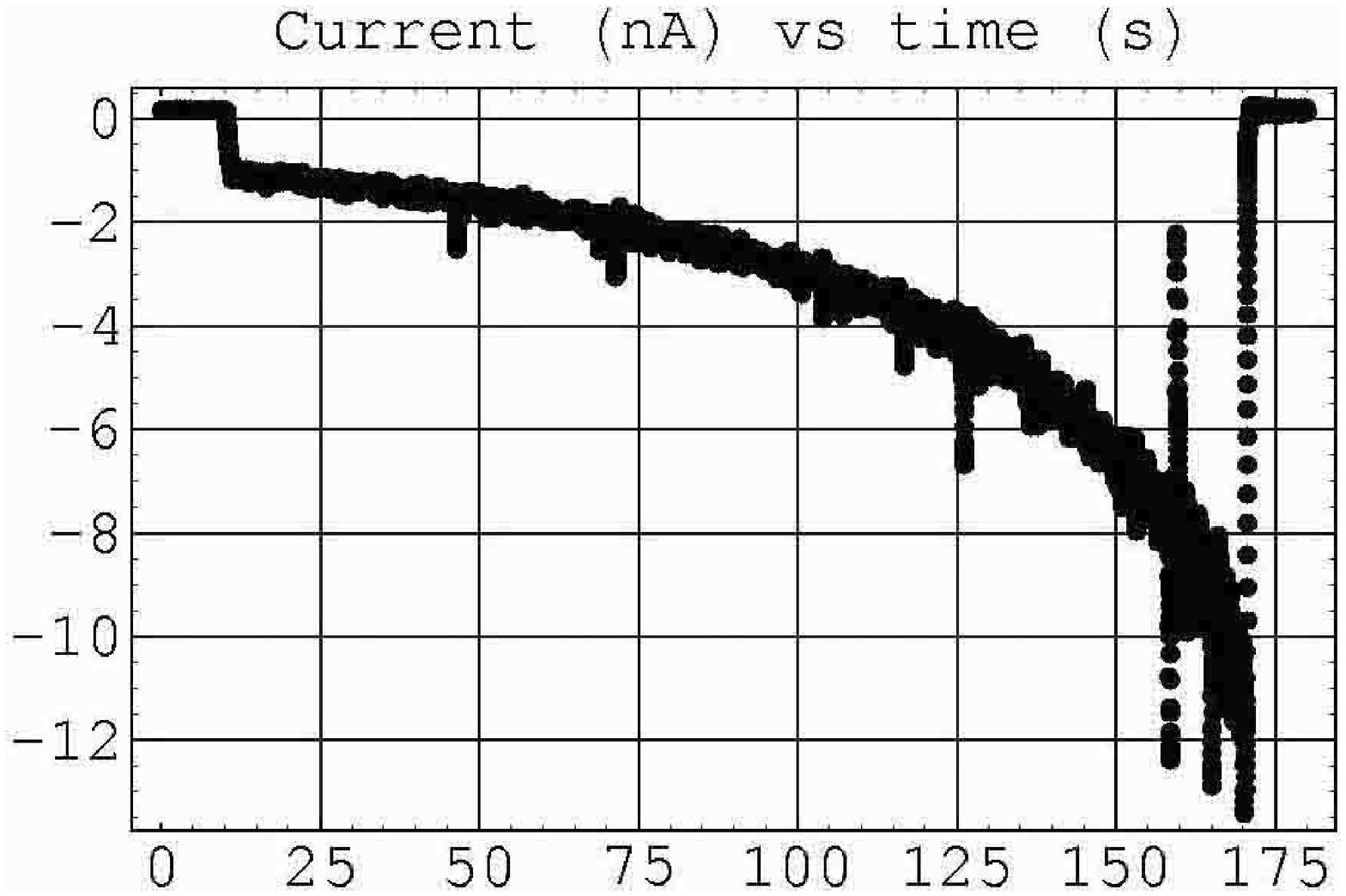}\\
\end{tabular}
\caption{\label{fig:idata} Output of
  the current amplifier in nA vs time in seconds for leakage current
  assessment.  In (a), the data are obtained with the LHe bath at 27.3 torr
  (2.05~K), an initial gap (visible in the transit) of 2.9~mm, and an initial
  potential of -6~kV at this gap.  The ground electrode is retracted
  at 5.08~cm~min$^{-1}$. Fig. (b) shows data from the return stroke of
  the ground electrode after a holding time of 16.8~min at the maximum
  gap of 7.8~cm; the LHe bath pressure is 30.0 torr (T = 2.09~K), and the
  electrode is
  returned at 5.08~cm~min$^{-1}$.  In Fig. (c), the LHe bath pressure is 
  600 torr (T = 3.98~K), the initial gap is 3.0~mm, the initial potential is
  29~kV, and the electrode is retracted at 2.54~cm~min$^{-1}$. Fig. (d) shows
  data from the return stroke of the ground electrode after a holding
  time of 11~hr and 32~min at the maximum
  gap of 7.2~cm; the LHe bath pressure is 600 torr (T = 3.98~K), and
  the electrode is
  returned at 2.54~cm~min$^{-1}$.}
\end{figure}

Stability of the voltage between the electrodes over time is
assessed by establishing the highest possible charge on the electrodes 
at maximum separation under the desired conditions, waiting for the
longest possible period under which the conditions can be maintained, 
and re-measuring the charge at the end of this period.  The initial
charge is measured in the usual way (Eq.~\ref{eq:charge}) during the outward
stroke of the ground electrode; the charge remaining after the holding
period is measured by applying Eq.~\ref{eq:charge} to the data
obtained as the ground electrode is returned to its 
original gap.  The difference in charge measured over the holding period is
then expressed as a leakage current.  Data used for leakage current
measurements in SF and normal state LHe are shown in
Fig.~\ref{fig:idata}

The leakage current from the HV electrode ($i_{HV}$) is determined from the
expression for the voltage, multiplied by the total capacitance of the
electrode at maximum spacing ($C_{HV}\approx C_{HG}+C_{HF}$ in
Fig.~\ref{fig:circuit}) and divided by the holding time
$\Delta t$:
\begin{equation}
i_{HV} = \frac{C_{HV}}{C_{HC}}\frac{\delta Q_{C}}{\Delta t}.
\label{eq:ileak}
\end{equation}
Here, $\delta Q_{C}$ is the {\it difference} in the charge
accumulation determined from
the data corresponding to the outward and inward strokes of the ground
electrode in Fig~\ref{fig:idata}.  From Figs.~\ref{fig:idata}a and b,
this difference as measured below the lambda point is consistent with
zero, when accounting for the uncertainties associated with amplifier zero
drift and transients.  The 95\% C. L. upper limit on the charge
difference is 14~nC.  The time $\Delta t$ between the outward and
inward strokes is 1009$\pm6$~s, limited by concern over the liquid
level dropping below the top edges of the electrodes after several
hours of tests.  From Figs.~\ref{fig:idata}c and d,
the charge difference as measured at atmospheric pressure is
also consistent with zero; the 95\% C. L. upper limit is 160~nC.  In this case 
the system had been topped off with LHe just before the first measurement 
and was left overnight, resulting in time $\Delta t = 41520\pm600$~s.
The electrode capacitance at maximum separation (again, 7.8~cm for the SF
data and 7.2~cm for the normal state) is found from the curve in
Fig.~\ref{fig:chv} to be $C_{HV} = 55 \pm 2$~pF and $57 \pm 2$~pF,
respectively.

Using the above measurements in Eq.~\ref{eq:ileak} results in 95\%
C. L. upper limit estimates of the leakage current
of 733~pA for the HV electrode in SF and
169~pA in normal state LHe.  While these results are
consistent with zero, the large uncertainty on the SF measurement
allows for values near the 1~nA design 
limit for the EDM experiment and is of some concern.  These
results are however, preliminary, and provision for several hours of
holding time below the lambda point will be made in subsequent tests
of the system. 

\subsection{Neutron Radiation Effects}
The Spallation Neutron Source will use a beam of high energy protons
(nominally 1~GeV with a 60~Hz repetition rate) directed at a liquid
mercury target to produce fast neutrons which will be partially
moderated by a liquid hydrogen cell.  The EDM experiment is scheduled to
be installed on a dedicated beam line with a neutron momentum centered
at 8.9~\AA ~(the optimum momentum for the production of UCN via
downscattering in SF LHe.)  While the series of
monochromators and frame definition choppers used to select the
momentum is expected to provide a narrow bandwidth, there
is some concern that performance of the HV and other systems in the
EDM experiment will degrade in the presence of a fast neutron background.
To check the operation of the HV test system under extreme background
conditions, charging and amplification tests are performed in the
presence of a 7~Ci PuBe neutron source.

The charging test (at atmospheric pressure) is performed in three
phases.  With the gap between the electrodes fixed at 3~mm, the HV
electrode is charged to the maximum possible potential in the absence of
the source.  The charging is repeated with the source placed next to
the exterior of the insulating vacuum system, behind a 2~cm thick slab
of polyethylene, at a point approximately 50~cm from the center of the 
front surface of the HV electrode.  Finally, the charging is repeated once more
with the polyethylene slab removed.  The maximum voltages attained,
together with the estimated neutron flux for each phase, are
shown in Table~\ref{tab:nrad}.  The results are consistent with a slight
improvement of the charging capability of the system in the presence
of an increasing fast neutron background.  However, the
test began less than 1~hr after a refill of the LHe volume and after a
few minutes of conditioning, so the slight improvement could
also be due to some extra conditioning.

With the source still in place, the system is again charged to 30~kV
and an amplification test is performed.  The electrodes are held at
maximum separation (7.2~cm), but only for a few minutes in the
interest of reducing
activation of the system components.  The voltage attained at this
separation is $475^{+38}_{-29}$~kV with no anomalous behavior observed,
consistent with the performance of the system at atmospheric
pressure in the absence of neutron radiation. 
\begin{table}
\caption{\label{tab:nrad}Neutron radiation effects on HV electrode
  charging.  The HV--ground gap is set to 3.0~mm, LHe bath pressure =
  600~torr (T = 3.98~K). Tests began $\sim$ 30~min after LHe fill,
  with $\sim$ 5~min prior conditioning.}
\begin{tabular}{|p{2.8cm}|r|p{2.8cm}|r|}\hline
\multicolumn{1}{|c|}{\it n-flux in gap}&\multicolumn{1}{|c|}{\it
  Breakdown~V (kV)}&\multicolumn{1}{|c|}{\it
  Comments}&\multicolumn{1}{|c|}{\it Time (min)}\\ \hline
(Background)&$30 \pm 1$&no source&0.0\\ \hline
$10^{6}$~s$^{-1}$, $E \sim 1$~MeV with ~10\% 1~keV admixture&$34 \pm
2$&source behind 2~cm polyethylene&3.0\\ \hline
$10^{6}$~s$^{-1}$, $E \sim 1$~MeV&$36 \pm 2$&polyethylene
  removed&7.0\\ \hline
\end{tabular}
\end{table}

For the amplification tests below the lambda point, the PuBe source is
placed in the same location but left on top of the polyethylene slab,
resulting in a neutron flux similar to that in the third row of
Table~\ref{tab:nrad}.  Tests at both positive and negative initial
voltages are performed; data corresponding to the outward stroke
of the ground electrode with positively charged HV electrode are shown
in Fig.~\ref{fig:transients}a.  Voltages obtained at maximum
electrode separation are $259^{+52}_{-16}$~kV and
$-274^{-55}_{+16}$~kV.  These values
are consistent with nominally better performance than in the absence
of the fast neutron background (Fig.~\ref{fig:hvdata}). However, the
pervasive transients observed in the data are a concern, and the large
uncertainty associated with them leaves the repeatability of these
results in question. 

\section{Conclusions and Outlook}
A prototype HV amplification system, consisting of a large, variable
parallel--plate capacitor has been constructed for a proposed neutron EDM
experiment in SF LHe.  The system has been used to amplify voltages in
the range of about 10~kV to 240~kV or greater across an electrode gap
of 7.8~cm in SF LHe.  This is roughly 60\% of the design goal for the
EDM experiment, which will have electrodes with a comparable gap, and
represents a lower limit on the dielectric breakdown strength of SF
LHe at large volumes.  These results are not appreciably changed in
the presence of a high background flux of fast neutrons.  Upper limits to
detectable leakage currents have been set at about 750~pA, nominally
tolerable in the final experiment.  Considerable improvement in
voltage tolerance and leakage current is observed with the system full of
normal state LHe at 4~K, performance that will likely be recoverable at lower
temperatures by pressurizing the LHe bath.  Additional statistics at
low temperatures with highly polished electrodes will be acquired to
establish a stable operating voltage, and to better understand the
conditions for bubble formation which is likely responsible for
measurement noise as well as reduced breakdown strength.    
Electric breakdown studies of candidate materials for the EDM
experimental test cells, using samples mounted behind the HV
electrode in the prototype system, are also anticipated in the near future.

\section*{Acknowledgments}
The investigators would like to thank V. Sandberg of Los Alamos
National Laboratory for suggesting the voltage amplification method,
and J. Jarmer of Los Alamos and J. Price of
the University of Colorado for advice on cooling the large LHe volumes
in the HV test system to below the lambda point.

\appendix
\section{The Effect of Transients}
\label{sec:transients}
\begin{figure}[htbp]
\begin{tabular}{p{7cm}p{7cm}}
\hspace{0cm} {\large (a)} & \hspace{0.5cm} {\large (b)} \\
\includegraphics[width=6.6cm]{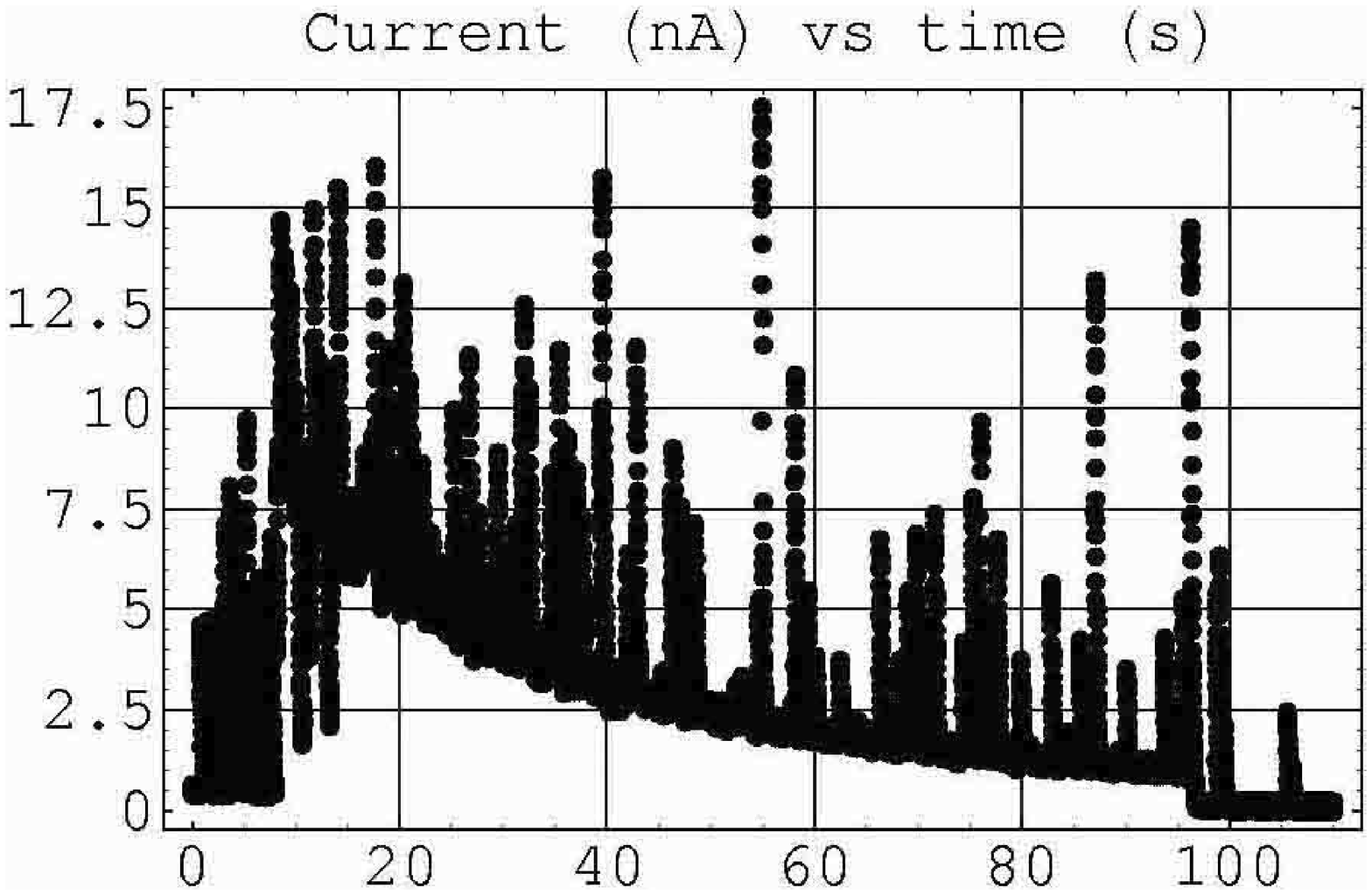} &
\includegraphics[width=6.6cm]{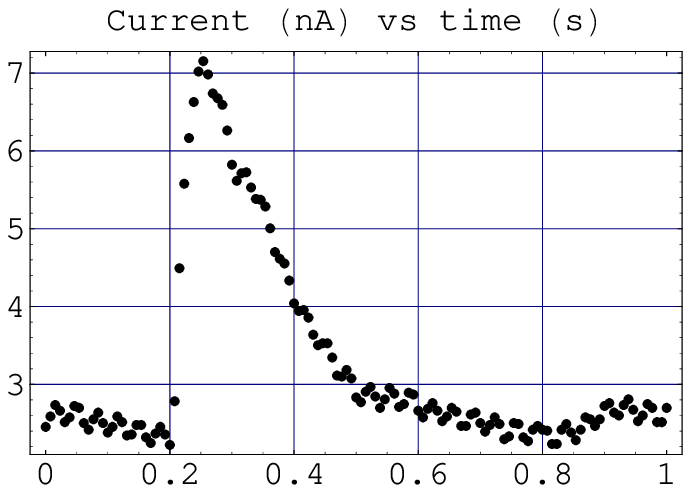}\\
\end{tabular}
\caption{\label{fig:transients} Output of
  the current amplifier in nA vs time in seconds illustrating
  transient effects.  Fig.~(a) shows an amplification test in the presence of
  a PuBe source.  The LHe bath pressure is 33.8 torr (T = 2.13~K), the
  initial gap is 2.8~mm, the initial potential is 13~kV, and the electrode
  retracted at 5.08~cm~min$^{-1}$.  Fig.~(b) is a 1~s time slice of
  an amplification data trace, showing a typical transient.}
\vspace{1cm}
\end{figure}

The transient current signals in the data traces 
(Figs.~\ref{fig:raw_data},~\ref{fig:idata},~\ref{fig:transients}) have
considerable influence on the uncertainty of the measured LHe properties.
This section summarizes the observed characteristics of the transients
and proposes a model for their explanation. 

An expanded view in time of a transient is shown in 
Fig.~\ref{fig:transients}b.  Typical transients have an amplitude of
about 1-25~nA. The usual rise time (10\%-90\%), FWHM, and decay times
are about 20~ms, 100~ms, and 150~ms, respectively.
Observations relating to transient formation include:
\begin{enumerate}
\item Transient signal shape (amplitude, rise and decay times, FWHM) does not
  change significantly with helium vapor pressure.    
\item Transients are almost exclusively positive (negative) when the HV
  electrode is positively (negatively) charged (compare
  Figs.~\ref{fig:raw_data}a and~\ref{fig:raw_data}c).
\item Transient density (number per unit time) increases at lower
  helium vapor pressures (compare
  Figs.~\ref{fig:raw_data}a and~\ref{fig:raw_data}b).
\item Transient density increases with applied field, at least at low
  vapor pressure (compare
  Figs.~\ref{fig:idata}a and b, and~\ref{fig:raw_data}a and c).
\item Transient density at low vapor pressure increases
  further in the presence of neutron radiation (compare
  Figs.~\ref{fig:raw_data}a and~\ref{fig:transients}a).
\end{enumerate}

Since the amplification signal is essentially DC, HV data are taken
with the input
filter on the current amplifier set to low-pass.  Specifically the
cutoff frequency is set to 3~Hz and the roll-off to 12~dB/octave.  To
simulate these conditions, tests are performed with a signal generator
connected to
the input of the amplifier, with the same filter settings, through a 
20~M$\Omega$ series resistor.  Output signals 
similar in rise time, FWHM, and decay time to the transients 
in the HV data can be obtained with square pulses 10~ms long and 
separated in time by a few hundred ms.  The pulse amplitude is attenuated by
about a factor of 5.  It is therefore probable that
the transient signal shape in the HV data is determined largely by
the amplifier input filter.  The actual transient phenomena are likely
significantly faster (at most a few ms) and larger (up to a few tens of nA).

The last three observations enumerated above suggest that transients
increase under conditions progressively more favorable for bubble
formation on the cathode electrode.  Bubbles with
radii on the order
of $r \sim$ 0.1--1~mm are observed to form on
the ground electrode during the initial charging of the system before
an amplification test.  They usually appear after a positive potential
of 10~kV or more has been established across the 2--3~mm gap, though detailed
observations of bubble formation and the conditions under which it
occurs or is enhanced have yet to be made.

The formation of bubbles in the gap can lead to a
coincident increase in the observed current at a particular gap
voltage.  This is because the dielectric constant of helium vapor or
vacuum ($\kappa = 1.00$) is lower than that of the normal state or SF
liquid ($\kappa = 1.05$); bubble formation leads to a decrease in the
HV--ground electrode capacitance.  To estimate this effect, the change
in capacitance $\Delta C$ due to the formation of a bubble of radius
$r$ is $\Delta C = 4\pi\epsilon_{0}r\Delta\kappa$.  For bubbles of
radius $r \sim$ 0.1--1~mm, $\Delta C =$ 0.5--5~fF.  A bubble
protruding 0.1--1~mm from the ground electrode in a typical field
between in the gap of 20~kV~cm$^{-1}$ sits at an
average potential of 200--2000~V.  While the time evolution of bubble
formation in the system has not been studied, the
formation of one of the larger bubbles on a time scale of $\Delta t \leq
1$~ms would lead to a current spike on the order of $V\Delta C/\Delta t
\geq 10$~nA, consistent with the amplitude of the observed
transients.  The sign of this current signal follows the sign of the
voltage, also consistent with observations.

While this model provides an ostensible explanation of the transients, one
drawback is that current signals of opposite sign, associated with the 
recovery of the capacitance once the bubbles detach from the
electrode and drift out of the gap, are expected to be
observed as well.  A lower limit on the average escape time of a
bubble from the gap can be estimated from the buoyant and drag forces
on the bubbles in the liquid.

Once detached from the electrode, bubbles float toward the surface of
the LHe under the influence of the buoyant force $F_{b} = V_{b}\rho_{l} g$,
where $V_{b}$ is the bubble volume, $\rho_{l}$ is the liquid density (assumed
constant through the bath depth at $1.3 \times 10^{2}$~kg m$^{-3}$)
and $g$ is the acceleration due
to gravity.  For bubbles with $r = 1.0$~mm, this force is initially
about 5~$\mu$N.

Motion of the bubble up through the liquid is opposed by a drag force
$F_{d}$, as well as an added--mass force $F_{m^{\prime}}$ induced by the
displacement of the fluid surrounding the bubble as it rises
\cite{Lamb45}.  For the case of spherical bubbles in a viscous fluid, 
\begin{equation}
F_{d}=12\pi\eta r {\bf v}; F_{m^{\prime}}=\frac{V_{b}\rho_{l}}{2}{\bf a}.
\end{equation}  
Here, $\eta = 1.6 \times 10^{-6}$~Pa~s is the viscosity of the normal-state
component of the LHe at 2.1~K, and {\bf v} and ${\bf a}$
are the bubble velocity and acceleration.  

Setting the buoyant force equal to the sum of the drag and added--mass
forces, and neglecting the bubble mass, the equation of motion for a
rising bubble reduces to:
\begin{equation}
{\bf a}-2g+\frac{18\eta}{r^{2}\rho_{l}}{\bf v}=0.
\label{eq:motion}
\end{equation}
Due to the low density of the LHe, the hydrostatic pressure at any
point along the electrode surfaces is determined almost exclusively by
the vapor pressure above the bath, even for the lowest vapor pressures
($\sim$ 30 torr) attained.  Therefore the radius of a rising bubble remains
essentially unchanged, and the coefficient of the ${\bf v}$ term is
constant.  For a bubble with $r = 1$~mm, the coefficient is about
0.1~s$^{-1}$, implying that a rising bubble does not reach an appreciable
fraction of its terminal velocity for $\sim 10$~s and accelerates
out of the gap with ${\bf a}\sim 2g$.  Assuming on average that
bubbles form near the middle of the electrode, bubbles have to
float up from rest through about 20~cm of LHe to escape the gap; the
average escape time is at least 100~ms.

The recovery of the capacitance presumably occurs as the bubble
drifts out though the fringe field near the rounded edge of the
electrode, over a time period shorter than but on the order of the escape
time.  If this interval is 10~ms or greater, the observed (opposite--sign)
current transient for a bubble escaping the gap would be at most about
1~nA, just above the resolution determined by the noise in
the readout (see, for example, the baseline in
Fig.~\ref{fig:transients}b).  This suggests that only the largest
bubbles escaping the gap would have a nominally visible effect, though the full
implication for the model as to whether bubbles must remain
attached to the electrodes (or must otherwise be prevented from a fast
escape from the gap) is unclear.

Nevertheless, if correct, the model suggests that the transients do not
correspond to an actual accumulation of charge during amplification
and can be removed from the data sets to assess the actual voltages
attained in the HV system.  However, the model is based only on a few
casual observations during initial operation of the system.
More careful studies of the actual signal profiles of the transients,
as well as of bubble formation, size, and number density as a function of time,
vapor pressure, applied voltage and electrode surface roughness will need to be
carried out in the near future.

\end{document}